\documentclass[prl, aps, 12pt, showpacs,epsfig]{revtex4}
\usepackage{graphicx}
\begin{document}

\title{Doubly Charged Lepton from an Exotic Doublet at the LHC}

\author{Teng Ma, Bin Zhang}
\email{zb@mail.tsinghua.edu.cn (Communication   author)}
 \affiliation{ Department of Physics,
Tsinghua University, Beijing, 100084, China\\
Center for
High Energy Physics, Tsinghua University, Beijing, 100084, China}
\author{Giacomo~Cacciapaglia}
\affiliation{Universit\'e de Lyon, F-69622 Lyon, France; Universit\'e Lyon 1, F-69622 Villeurbanne;
CNRS/IN2P3, UMR5822, Institut de Physique Nucl\'aire de Lyon,
F-69622 Villeurbanne Cedex, France}

\begin{abstract}
     We studied the signatures at the LHC of the electroweak SU(2) lepton doublet which embeds a doubly charged lepton. The doubly charged lepton pair and single production rates, which are different from the triplet case, are studied and all the detectable decay modes are considered. We also applied the same kinematic cuts as in triplet case to reduce the background at LHC and the needed integrated luminosities to find the doubly charged leptons with different masses are analyzed. We also suggested a method to distinguish the exotic lepton doublet and triplet.

\end{abstract}

\pacs{14.60.Hi, 14.60.Pq, 14.60.St, 12.15.Ff}

 \maketitle
\section{introduction}

After the discovery of the Higgs boson, the main goal of the LHC experiments is to discover signs of new physics.
Searches based on the most popular models, like supersymmetry, in their incarnations that lead to easy-to-see signatures have lead so far to stronger and stronger bounds, without any sign of departure from the Standard Model predictions.
In 2015, the LHC will end the two year technical pause and be restarted at a higher center of mass energy close to 14 TeV. The higher energy will allow the experiments to access higher mass scales, but also electroweak processes that gave too feeble cross sections to be detected during the 8 TeV run.

In this paper, we will focus on one such case, where the new particles are new vector-like leptons, whose mass is not generated by the Brout-Englert-Higgs mechanism.
New particles of this kind appear in a few models of new Physics, like models of TeV-scale see-saw neutrino masses~\cite{Franceschini:2008pz,delAguila:2008cj,delAguila:2008hw,Arhrib:2009mz,McDonald:2013hsa}, little Higgs~\cite{Cacciapaglia:2009cv}, two Higgs doublets ~\cite{Alves:2013dga}, minimal technicolor~\cite{Frandsen:2009fs}.
The new leptons are typically produced at the LHC via their electroweak couplings to the $W$ and $Z$ bosons, and the photon, and their decays depend crucially on how they couple to the standard leptons.
We can classify the phenomenology in three separate classes: models where the mixing takes place via Yukawa couplings to the standard chiral leptons via the Higgs doublet~\cite{delAguila:2008pw,Delgado:2011iz,Ma:2013tda,Altmannshofer:2013zba}; models where they couple via higher order operators like four-fermion interactions~\cite{Alloul:2013raa}; finally models where they are odd under a certain parity and couple to standard leptons via a Dark Matter candidate, like in little Higgs models with T-parity~\cite{Cacciapaglia:2009cv}.
Lepton-like states, i.e. new vector-like fermions that do not carry color, can also be used to contain a fermionic Dark Matter candidate if the new representation does not mix with the standard leptons~\cite{Cirelli:2005uq}.

We are interested here in the case of mixing via a Yukawa interaction
to the Higgs, and in particular in cases where the new leptons contain
a doubly-charged state.
The mixing via the Higgs arises typically in models addressing the
shortcomings of the Higgs scalar field in the standard model, like for
instance in models in warped space~\cite{Csaki:2008qq} or models of
compositeness~\cite{delAguila:2010vg}. In the latter scenario, doubly
charged leptons naturally arise once extended custodial
symmetries~\cite{delAguila:2010es} are summoned to protect the
couplings of the standard leptons to the $Z$ boson.
Phenomenologically, the presence of a doubly charged lepton offers a
golden channel at the LHC, because it can only decay to a $W$ boson
and a charged lepton carrying the same charge, thus leading to a
relatively clean same-sign di-lepton final state.
As the quantum numbers of the new leptons are limited by the
requirement of a Yukawa coupling via a Higgs
doublet~\cite{delAguila:2008pw}, a doubly charged state can occur only
in two cases: if the new leptons belong to a triplet of SU(2) with
hypercharge $-1$, or a doublet with hypercharge $-3/2$.
The former case have been extensively studied
in~\cite{Delgado:2011iz}, thus we will here focus on the doublet which
is predicted, for instance, in the model of
Ref.~\cite{delAguila:2010es}.
After introducing the general parametrization of a model that is a
simple extension of the standard model plus the new lepton doublet, we
will present the results of the application of the searches studied in
our previous work~\cite{Ma:2013tda} to the doublet case, and discuss
the prospects to distinguish the two representations at the LHC at 14
TeV.

\section{the model}

We consider an effective model consisting of the Standard Model complemented with a new exotic un-colored doublet $X$ with hypercharge $Y_X = -3/2$:
\begin{equation}
X = \left( \begin{array}{c}
X^- \\ X^{--} \end{array} \right) \in (2,-3/2)\,,
\end{equation}
and we further assume that the doublet has both chiralities transforming in the same way.
The most general lagrangian will therefore contain a gauge invariant mass term $M_1$, and new Yukawa couplings to the standard model leptons
\begin{equation}
- {\cal L}_{\rm Yukawa/mass} = \lambda^{ij}_1\, \overline{L}^i_L H {e^j_R} + \lambda_2^{j} \,\overline{X}_L H^c {e^j _R} + M_1\,  \overline{X}_L X_R + h.c.\,,
\end{equation}
where the flavor indices $i,j$ span over the 3 standard generations, and $L/R$ label the chirality of the fermion.
We can now use the flavor symmetry in the standard chiral leptons to diagonalize the standard Yukawa matrix $\lambda_1^{ij}$, while an arbitrary phase redefinition of the new doublet components allows us to reabsorb the phase of $M_1$ and one of the three $\lambda_2^j$.
After the Brout-Englert-Higgs doublet $H$ develops its vacuum expectation value, the new Yukawa couplings induce a mixing between the new doublet and the standard chiral fermions:
\begin{eqnarray}
 - {\cal L}_{\rm mass} &=& m^i_1\, \overline{e}^i_L {e^i_R}  + {m_2}^j \, \overline{X}_L e^j_R +  M_1\, (\overline{X}^{--}_L X^{--}_R + \overline{X}^{-}_L X^{-}_R) + h.c.\,,
\end{eqnarray}
where the flavour index now labels $i=e, \mu, \tau$.
In the charged lepton sector, the mass matrix is very similar to the one obtained in the triplet model~\cite{Delgado:2011iz,Ma:2013tda}, the only difference being that the chiralities in the new Yukawa coupling are inverted: in this case, it is the left-handed component of $X^-$ that mixes with the standard leptons.
The main consequence of this is that the mixing in the right-handed sector is dominant, while the mixing in the left-handed one is very small being suppressed by the small masses of the standard leptons.
Following the notation in~\cite{Delgado:2011iz} we can define mixing matrices in the left-handed sector $S_E$ and right-handed one $T_E$ that diagonalize the mass matrix in the charged lepton sector.
As in the triplet case, the couplings of the new exotic leptons to the standard ones depend only on 3 elements of the right-handed matrix, so we define
\begin{equation}
v^i = T_E^{\ast,4i}\,, \quad i = \mbox{e, $\mu$, $\tau$}\,.
\end{equation}
The couplings of the exotic states to the light leptons can therefore be written as:
\begin{eqnarray}
g_{Z} (Z_\mu \bar{e}^j_R X^-_R) &=& \frac{g}{2 \cos \theta_W} v^j \,, \nonumber \\
g_{H} (H \bar{e}^j_R X^-_L) &=& - \frac{M_1}{v} v^j\,, \nonumber \\
g_{W} (W^+_\mu \bar{e}^j_R X^{--}_R) &=& \frac{g}{\sqrt{2}} v^j\,.
\end{eqnarray}
We have here neglected the couplings suppressed by the light lepton masses: note the absence of a coupling of $X^-$ to neutrinos via the $W$, which is left-handed and thus suppressed.
This implies a first important difference with respect to the triplet case: the singly-charged exotic lepton mainly decays into neutral boson plus a charged lepton.

In the limit of vanishing masses for the light leptons, i.e. $m_1^i = 0$, the mass eigenvalues are
\begin{equation}
m_{X^{--}}^2 = M_1^2\,, \qquad m_{X^-}^2 = M_2^2 = M_1^2 + \sum_{i=e,\mu,\tau} |m_2^i|^2\,;
\end{equation}
while the relevant mixing angles are given by
\begin{equation}
v^i = - \frac{m_2^i}{M_2}\,.
\end{equation}
Defining
 \begin{equation}
\lambda = \sum_{i=e,\mu,\tau} |v^i|^2,  \nonumber
\end{equation}
the two masses are related by
\begin{equation}
M_2^2 = \frac{1}{1-\lambda} M_1^2\,,
\end{equation}
therefore $M_2 > M_1$\,.
Adding the light lepton masses amount to a subleading correction to the above formulae, as shown in~\cite{Ma:2013tda}.

The values of $v_i$ are constrained by corrections to the couplings of the standard leptons: while lepton changing processes give very strong bounds on the product of two elements, the only absolute bounds on each $v_i$ is given by the corrections to the couplings of the light leptons to the $Z$ boson, which have been precisely measured at LEP~\cite{LEP}.
In the doublet case, the corrections are dominantly right handed and we can put bounds by conservatively compare the correction to the coupling due to the mixing with the error in the measurement.
Neglecting the correlation between the measurements in different flavors, and at 3$\sigma$ we have
\begin{equation}
\left| \frac{\delta g_{Z l^+_R l^-_R}}{g_{Z}^{\rm SM}} \right| = \frac{|v^i|^2}{2 \sin^2 \theta_W} < \left\{ \begin{array}{l}
0.38 \% \\ 1.7 \% \\ 0.8 \%
\end{array} \right. \quad \mbox{for} \quad \begin{array}{c}
e \\ \mu \\ \tau
\end{array} \quad \Rightarrow \begin{array}{l}
|v_e|^2 < 1.7 \cdot 10^{-3} \\
|v_\mu|^2 < 7.5 \cdot 10^{-3} \\
|v_\tau|^2 < 3.6 \cdot 10^{-3}
\end{array}
\end{equation}
Like for the triplet case, the mass splitting between the singly and doubly charged exotic leptons is therefore bound to be small, even when electroweak loop corrections are included.
Unless $\lambda < 10^{-12}$~\cite{Delgado:2011iz}, therefore, the exotic leptons will predominantly decay into a standard lepton plus a boson:
\begin{equation}
X^- \to l^- (Z, H)\,, \qquad X^{--} \to l^- W^-\,.
\end{equation}
The total decay widths of the exotic doublet leptons are
\begin{eqnarray}
&&\Gamma_{X^{--}} = \frac{G_{F}M^3}{8\sqrt{2}\pi}F_{1}(r_{W})\lambda, \nonumber\\
&&\Gamma_{X^{-}} = \frac{G_{F}M^3}{16\sqrt{2}\pi}(F_{1}(r_{Z})+F_{0}(r_{H}))\lambda;
\end{eqnarray}
where
\begin{equation}
 F_{n}(r)=(1+2nr)(1-r)^2,
 \end{equation}
and $r_{a}=m_{a}^2/M_{X}^2$.

\begin{figure}
\begin{center}
\includegraphics[width=0.45\textwidth]{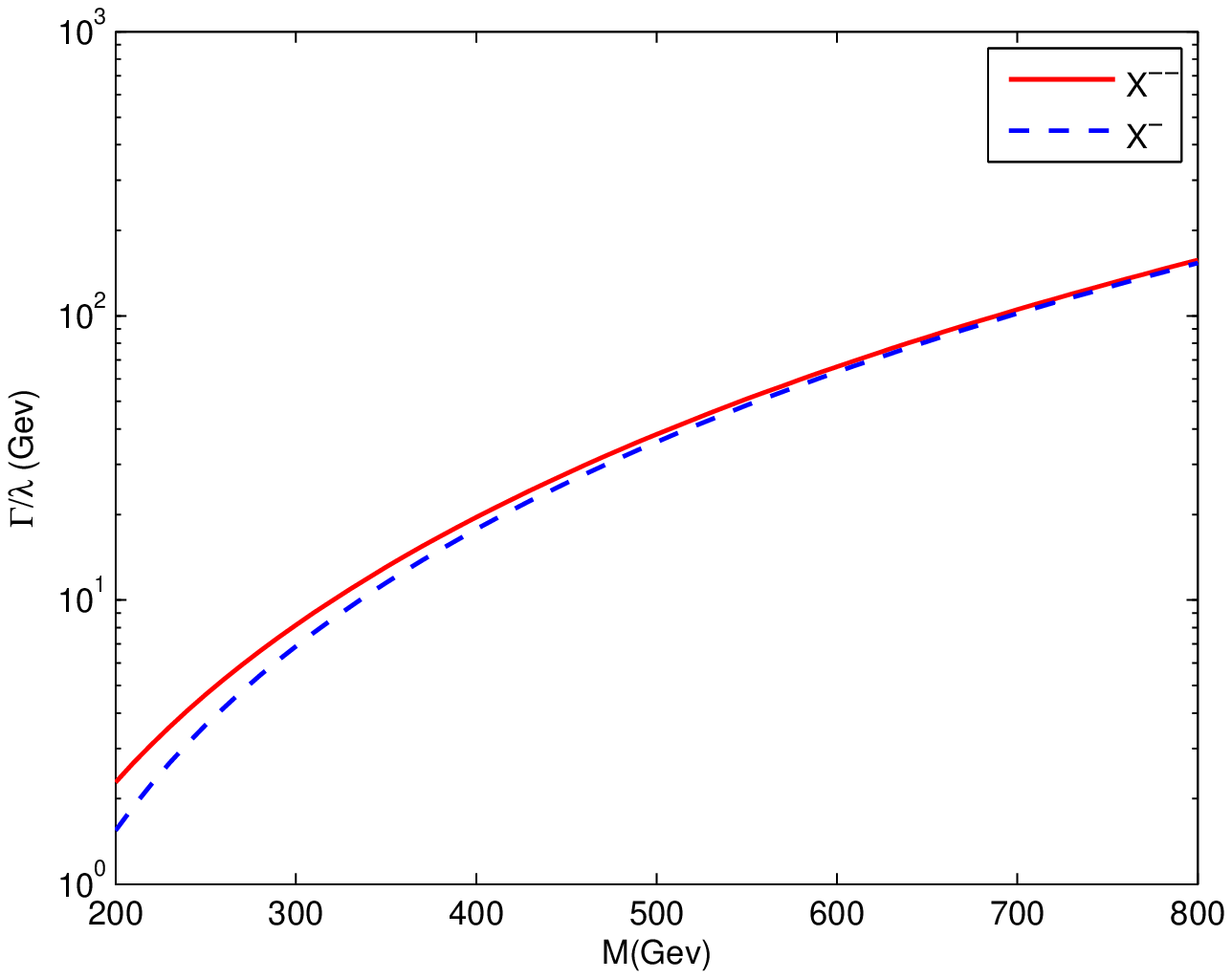}
\includegraphics[width=0.45\textwidth]{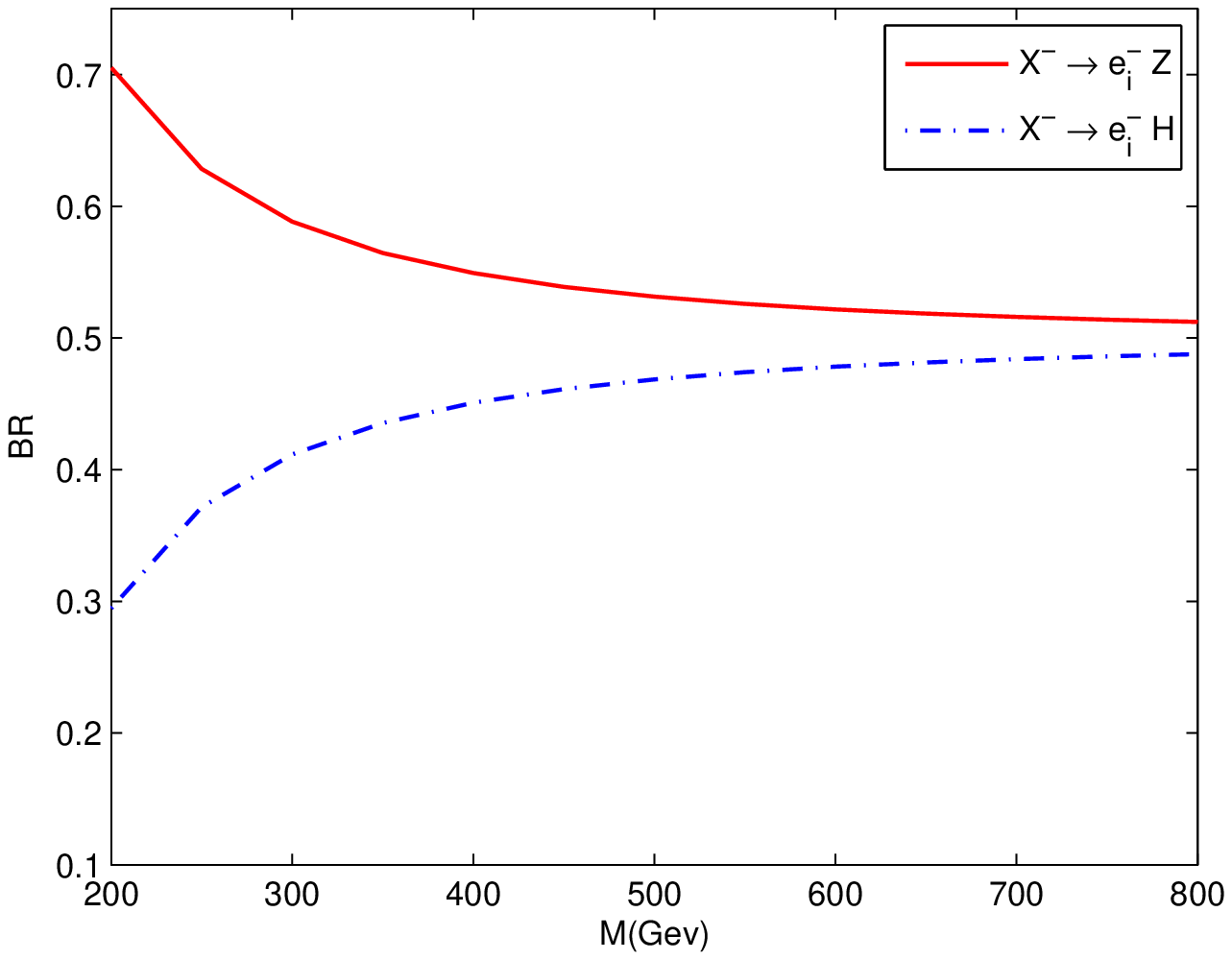}
\caption{Total decay widths in unit of $\lambda$ and branching ratios of the singly charged lepton as a function of the mass.}\label{figure-Doubletdecay}
\end{center}
\end{figure}

In the following we will work in the region where the main decays are 2-body decays into a standard boson, i.e. $\lambda > 10^{-12}$.
In this range, the total decay widths of the doublet lepton in units of $\lambda$ are shown in Fig.\ref{figure-Doubletdecay}, together with the Branching Ratios for the singly charged component.
While $X^{--}$ always decays into a $W$ boson, like in the triplet case, the singly charged $X^-$ only decays into neutral bosons $Z$ or $H$.
The rates into the various lepton flavors depend on the ratios between the various mixing elements: in the numerical results we will assume democratic decays, i.e. $|v_e|^2=|v_\mu|^2=|v_\tau|^2$ as a basic sample. However, the results can be generalized by simple rescaling. For future reference, we define the rate of decays into taus as
\begin{equation}
\zeta_\tau = \frac{|v_\tau|^2}{\lambda}\,.
\end{equation}

\begin{figure}
\begin{center}
\includegraphics[width=0.6\textwidth]{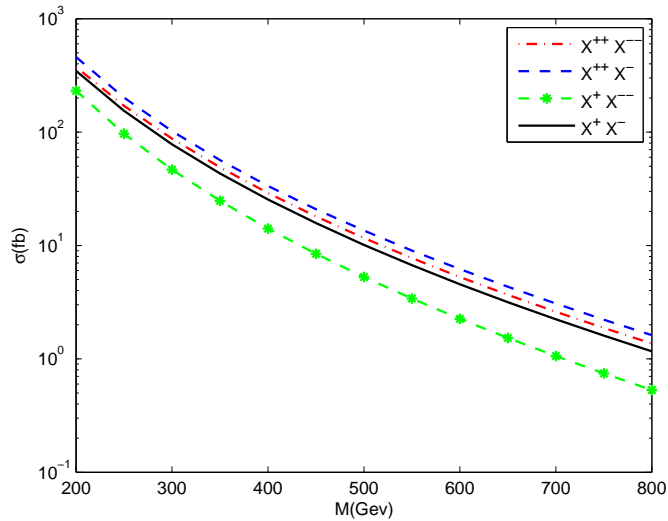}
\caption{Production cross sections for a pair of exotic doublet leptons as a function of the mass at LHC 14 TeV.}\label{figure-Doubletproduction}
\end{center}
\end{figure}

Finally, in Figure~\ref{figure-Doubletproduction} we show the cross sections for the production of two exotic leptons belonging to the doublet.
It is instructive to compare this figure with the corresponding one for the triplet in~\cite{Delgado:2011iz,Ma:2013tda}: we see that the production channels of the doubly charged lepton ($X^{--} X^{++}$, $X^{++} X^-$ and $X^{--} X^+$) are all reduced by a factor of a few, mainly de to the smaller couplings of the doublet to SU(2) gauge bosons with respect to the triplet.
On the other hand, the pair production of singly charged leptons, $X^+ X^-$, is enhanced in the doublet case due to its larger couplings to the $Z$ boson: this fact will play a crucial role in the possibility to distinguish the two cases, triplet and doublet, at the LHC.

\section{Phenomenology of the Lepton Doublet}

In this section we will apply the same searches that we designed for the triplet~\cite{Ma:2013tda}, to the exotic doublet case.
We will study the following 3 processes:

\begin{eqnarray}
\mbox{A)} &\Rightarrow&  pp\rightarrow  \left\{ \begin{array}{l}
X^{--}X^{++}\rightarrow \ell^- W^-\ell^+W^+ \\
X^{--}X^{+} +h.c.\rightarrow \ell^- W^-\ell^+Z + h.c. \end{array} \right\} \rightarrow \ell^- \ell^+ jj \ell^-\bar{\nu}(\ell^+\nu)
 \\
\mbox{B)} &\Rightarrow& pp\rightarrow  \left\{ \begin{array}{l}
X^{--}X^{+} +h.c.\rightarrow  \ell^- W^-\ell^+Z + h.c.  \\
X^{+}X^{-} \rightarrow \ell^- Z \ell^+Z  \end{array} \right\} \rightarrow \ell^- \ell^-Z(\ell^+ \ell^+) jj
 \\
\mbox{C)} &\Rightarrow& pp\rightarrow \left\{ \begin{array}{l}
 X^{--}X^{+} +h.c.\rightarrow  \ell^- W^-\ell^+Z + h.c.  \\
X^{+}X^{-} \rightarrow \ell^- Z \ell^+Z + h.c. \end{array} \right\} \rightarrow \ell^- \ell^- Z(b\bar{b}) jj
 \\
\mbox{D)} &\Rightarrow& pp\rightarrow  \left\{ \begin{array}{l}
X^{--}X^{+} +h.c.\rightarrow  \ell^- W^-\ell^+H + h.c. \\
X^{+}X^{-} \rightarrow \ell^- Z \ell^+H + h.c. \end{array} \right\} \rightarrow \ell^- \ell^- H(b\bar{b}) jj
 \\
\mbox{E)} &\Rightarrow& pp\rightarrow X^{+}X^{-} \rightarrow \left\{ \begin{array}{c}
l^+ Z l^- Z \\
l^+ Z l^- H + h.c. \\
l^+ H l^- H
\end{array} \right\} \rightarrow l^+ l^- b \bar{b} b \bar{b}
\end{eqnarray}
Here a lepton is either an electron or muon, and we leave the more difficult study of taus to a further study.
The analysis for the case $A$ ($\ell^- \ell^+ jj \ell^-\bar{\nu}(\ell^+\nu)$) has been first proposed in~\cite{Delgado:2011iz}.

We generated the signal events with MadGraph5~\cite{MadGraph}, and used the same background samples we used in our previous work~\cite{Ma:2013tda}.
For all analyses, the events are generated with the following basic cuts on the leptons and jets:
\begin{eqnarray}
&&p_T(\ell) > 15 {\,\rm GeV},\,|\eta(\ell) | < 2.5,\nonumber\\
&&p_T(j) > 15  {\,\rm GeV},\,|\eta(j) | < 2.5,\nonumber\\
&&\Delta R(ll) > 0.3,\,\Delta R(jl) > 0.4,\,\Delta R(jj) > 0.4. \label{3}
\end{eqnarray}
Furthermore, in our numerical results we assume that the exotic leptons decay democratically into the 3 families of leptons, with a Branching Ratio of $\frac{1}{3}$.
However, as the signal rescales with the Branching Ratios, our results can be generalized to a more generic scenario by rescaling the signal event rates in all channels by a factor
\begin{equation}
\frac{9}{4} (1-\zeta_\tau)^2\,.
\end{equation}

\subsection{Doubly charged lepton pair production: channel A}

\begin{figure}
\begin{center}
\includegraphics[width=0.6\textwidth]{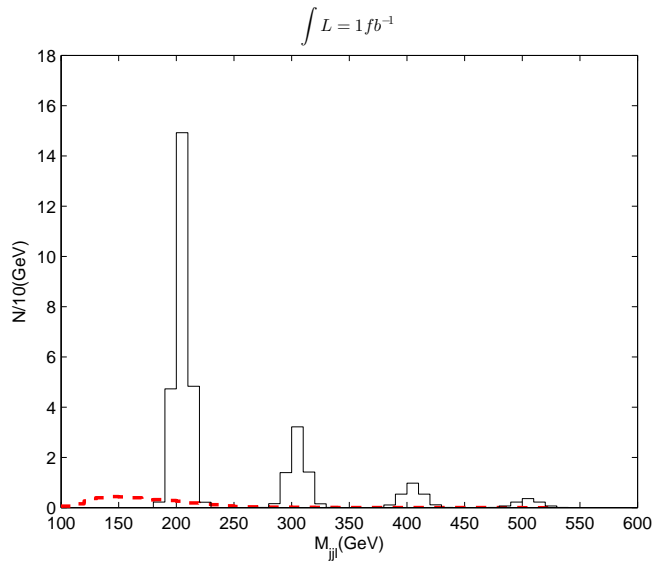}\\
\caption{Invariant mass $M(\ell jj)$ distribution curves of exotic leptons signal resonances and the background.}\label{figure77-4}
\end{center}
\end{figure}

The pair production of the doubly-charged lepton offers the ideal channel for the discovery of the new states: in fact, when one of the two $W$'s in the final state decay leptonically, one can reconstruct the mass of one lepton via the two same-sign leptons and the missing energy, and compare with the mass reconstructed from the third lepton and the hadronically decaying $W$.
The same final state is however obtained from the associated production of the doubly-charged lepton together with its singly-charged partner: in this case, the two jets can come from the $Z$ boson.
In order to add the latter contribution to the signal, we define a cut on the invariant mass of the two jets that includes both $W$ and $Z$ resonances:
\begin{equation}
 M_W-20\,{\rm GeV} <M(jj) <M_Z + 20 \,\rm GeV, \qquad \mbox{and} \quad  E\!\!\!\slash_T>25 {\,\rm GeV}.\label{4}
\end{equation}
We can then further reduce the background by requiring the mass reconstruction of the two exotic leptons
\begin{equation}
| M(jj\ell^\pm) - M( \ell^\mp \ell^\mp \bar{\nu}) | < 30 \,\rm GeV,
\end{equation}
where we rely on the fact that the masses of the two exotic leptons are very close, as discussed in the previous section.
The two mass cuts are very effective in reducing the standard model background, as shown in Figure~\ref{figure77-4} where we plot the distribution of invariant mass $M(\ell jj)$ for the background and signal for various masses of the exotic leptons.
We see that the number of background events around the mass of the exotic lepton is always negligible with respect to the signal events.

\begin{figure}[tb]
\begin{center}
\includegraphics[width=0.45\textwidth]{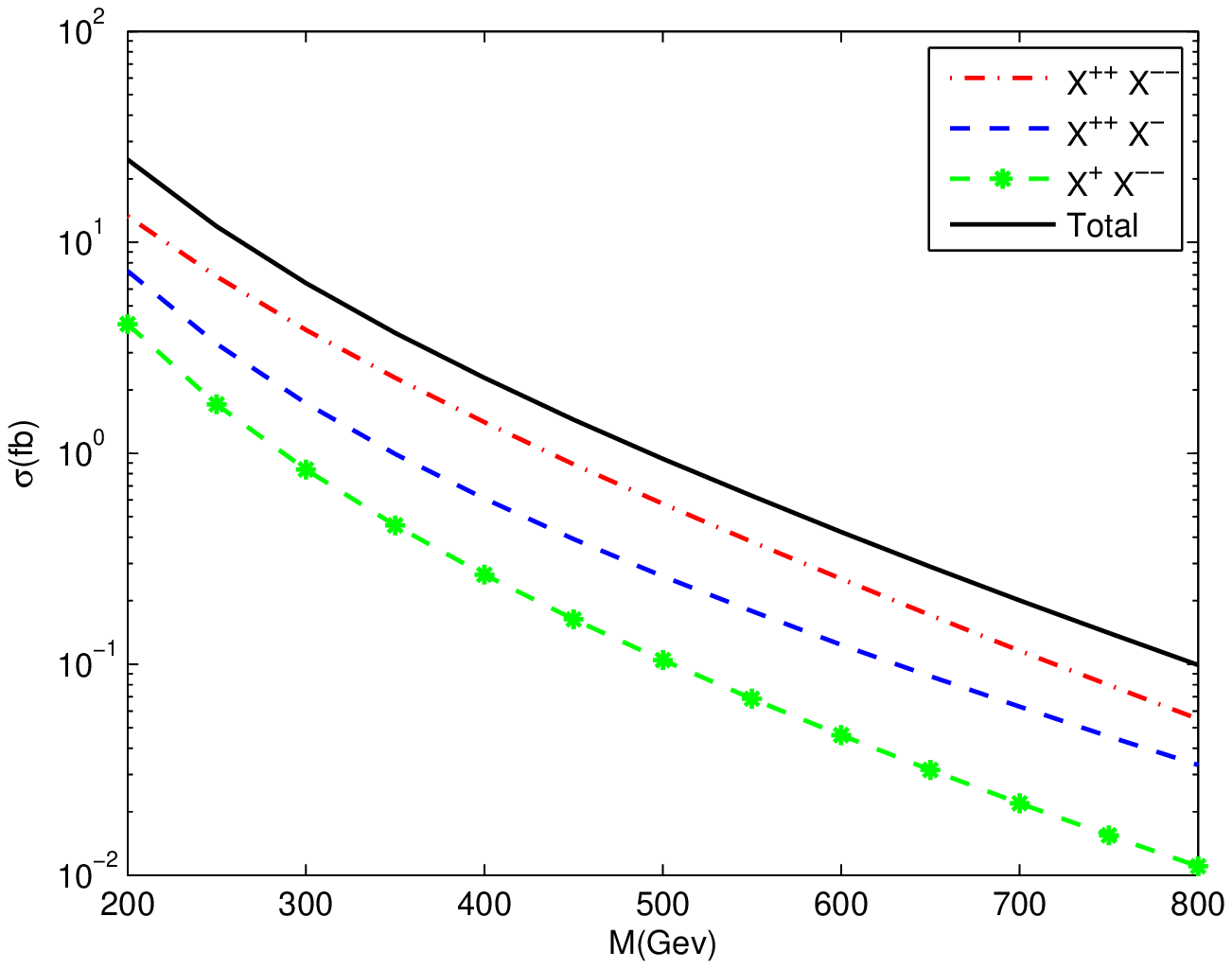}
\includegraphics[width=0.45\textwidth]{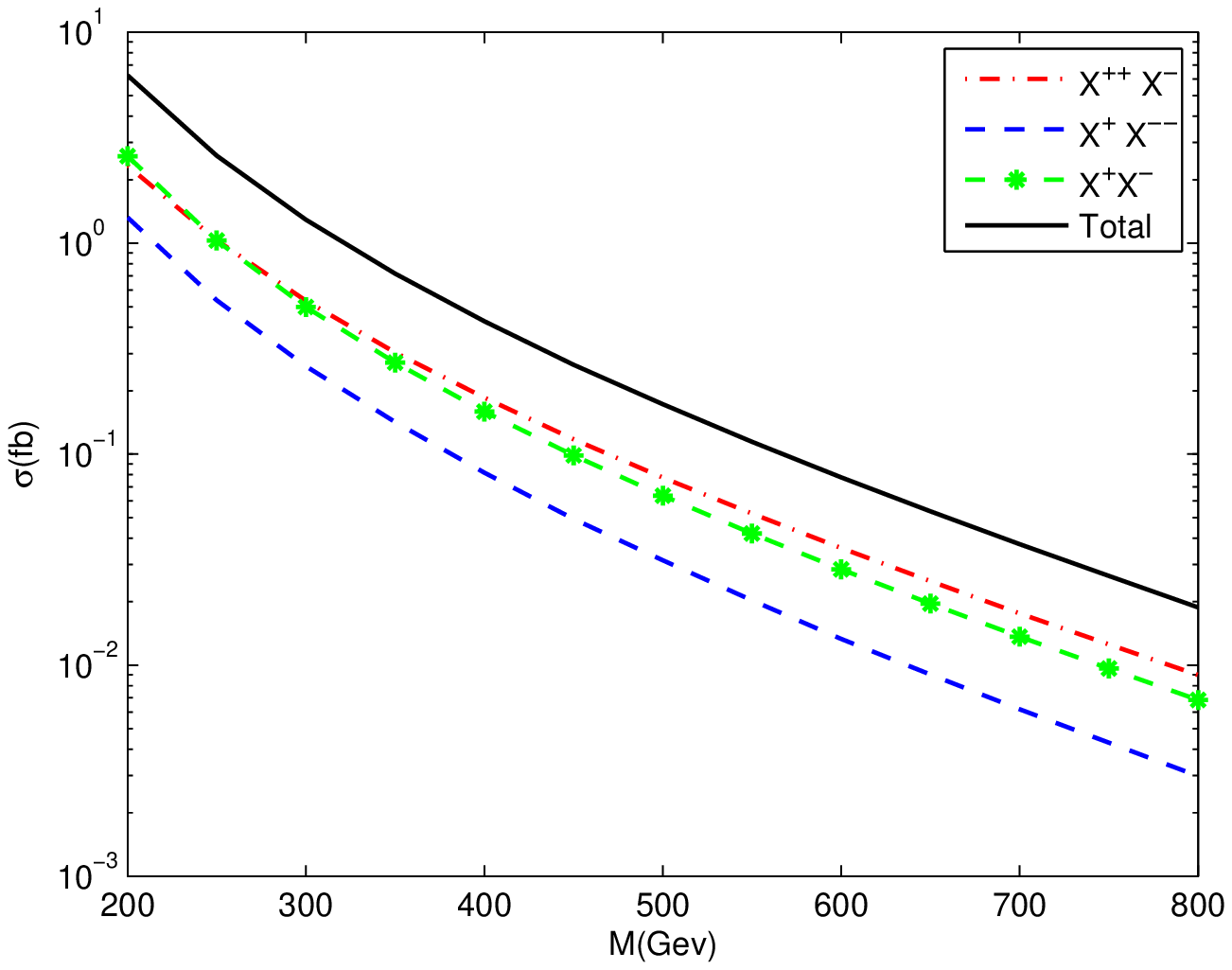}
\includegraphics[width=0.45\textwidth]{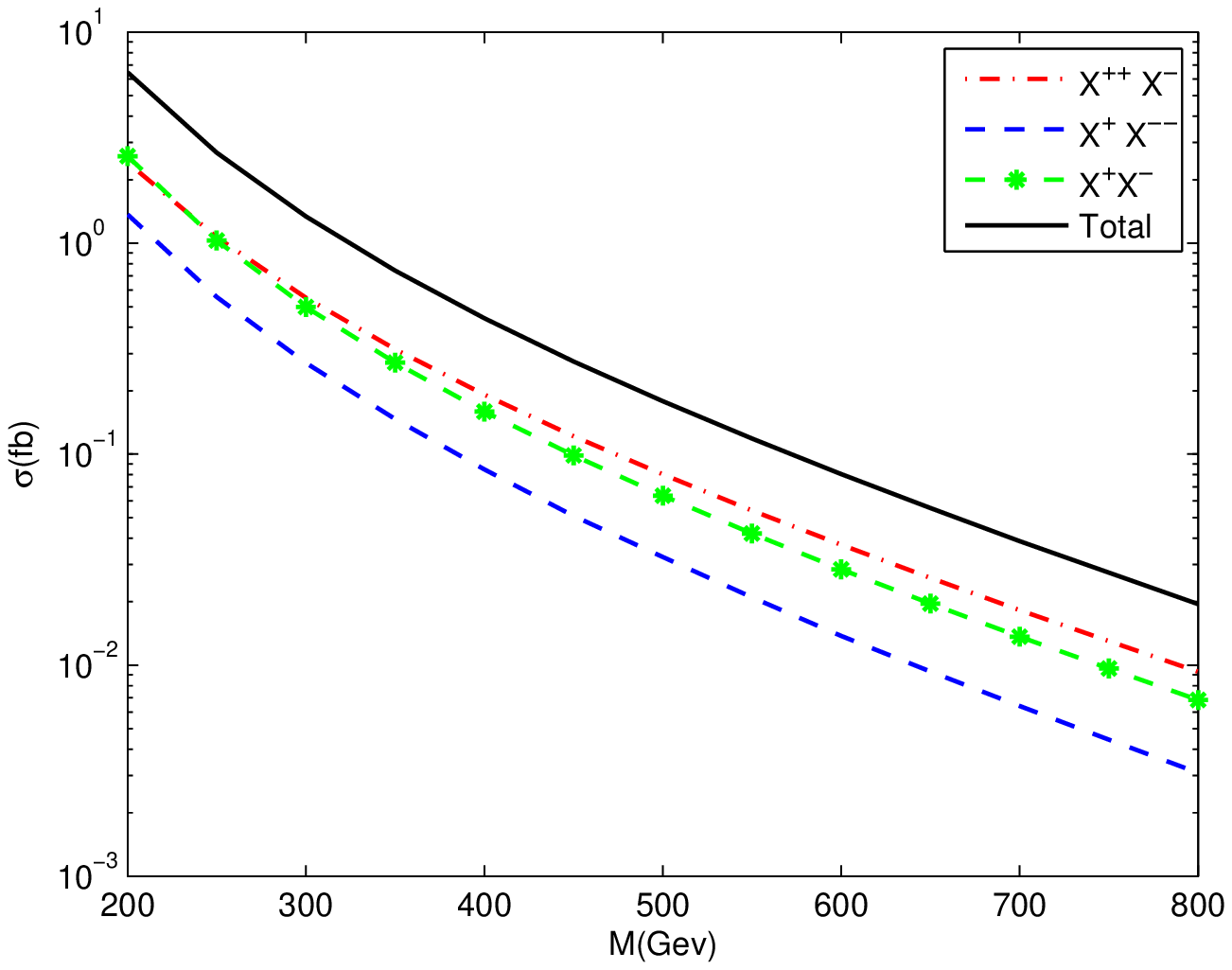}
\includegraphics[width=0.45\textwidth]{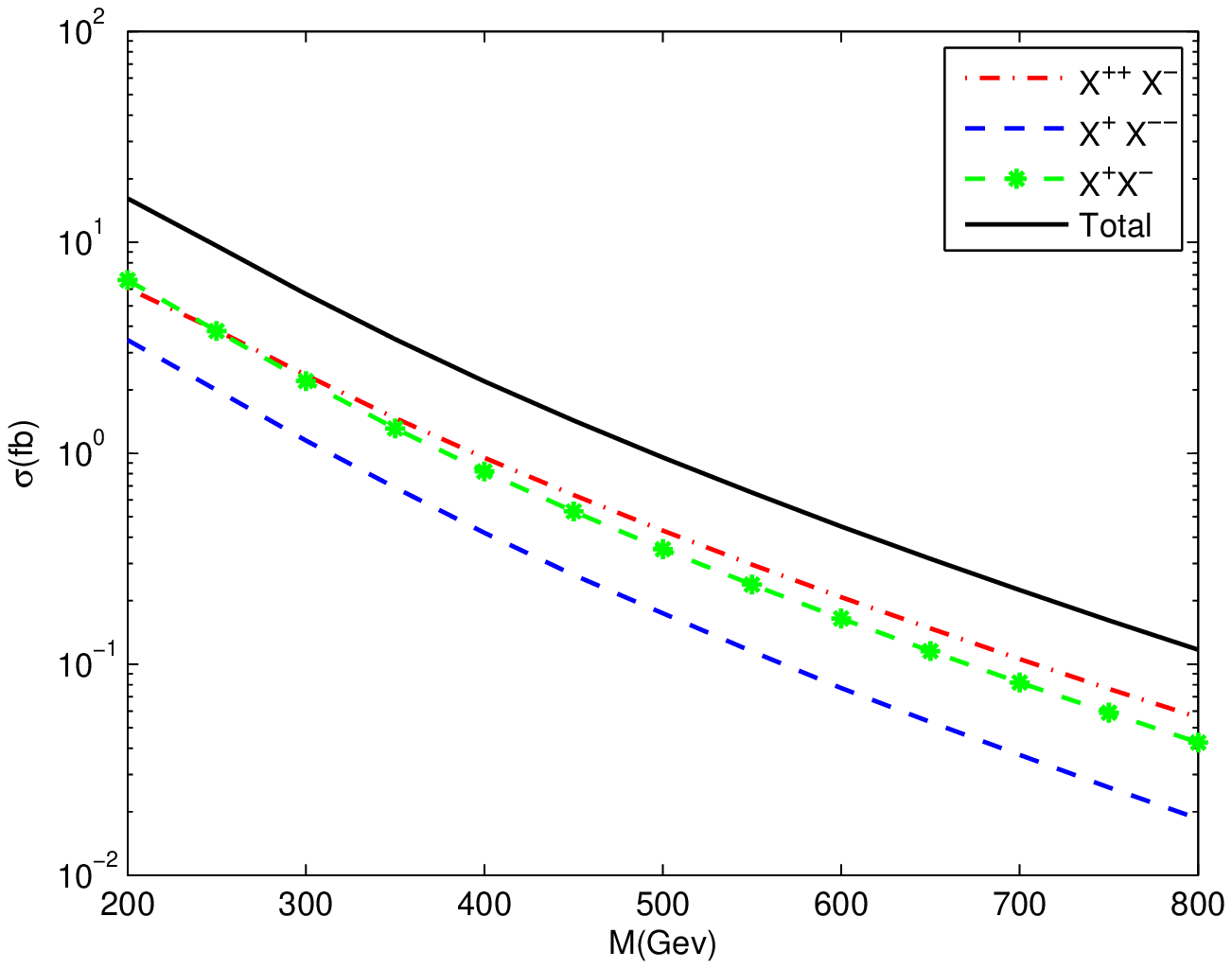}
\caption{The final lepton doublet signal production rate for the channels A (top-left), B (top-right), C (bottom-left) and D (bottom-right) at 14 TeV LHC.}\label{figure50-1}
\end{center}
\end{figure}

\begin{figure}[tb]
\begin{center}
\includegraphics[width=0.45\textwidth]{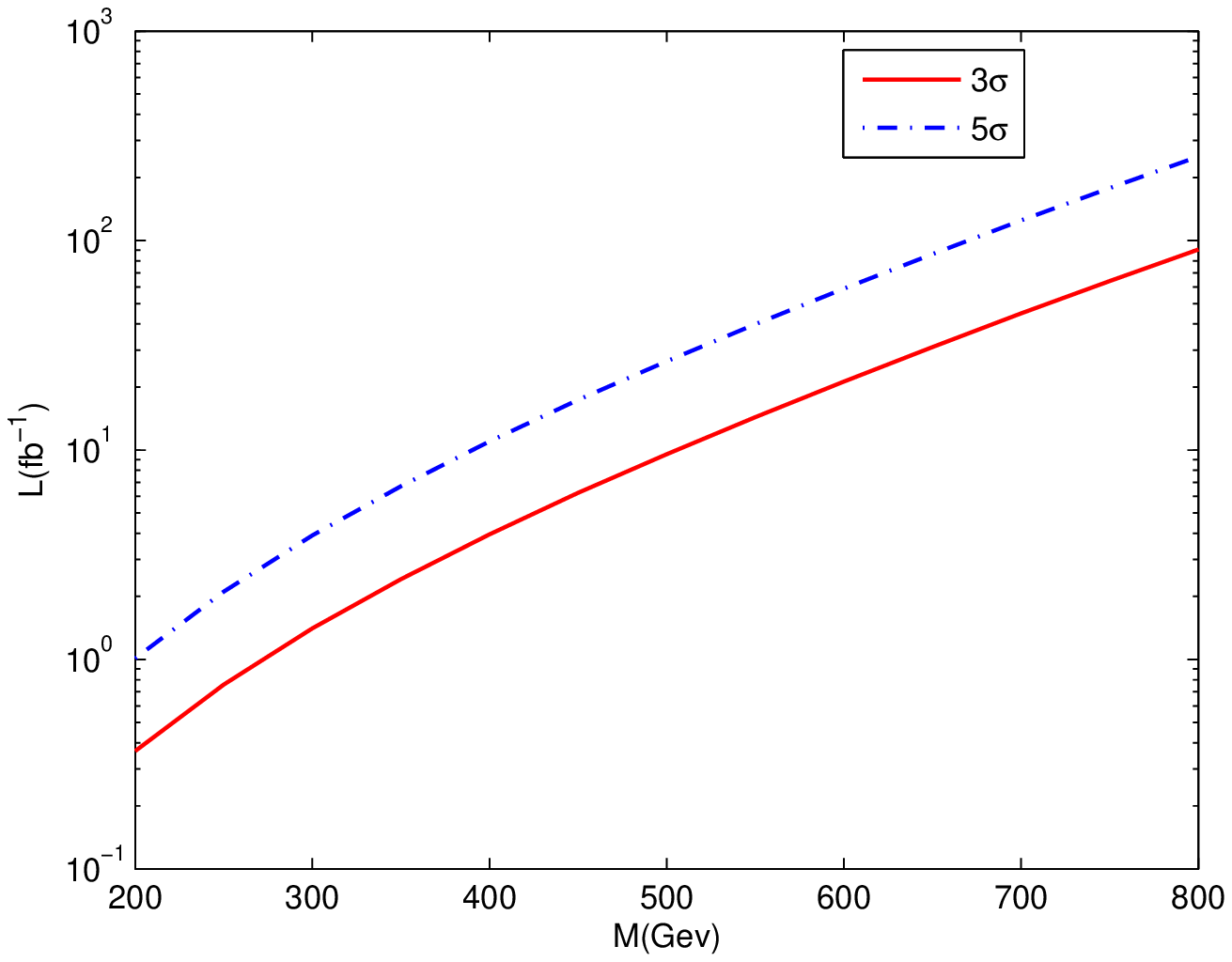}
\includegraphics[width=0.45\textwidth]{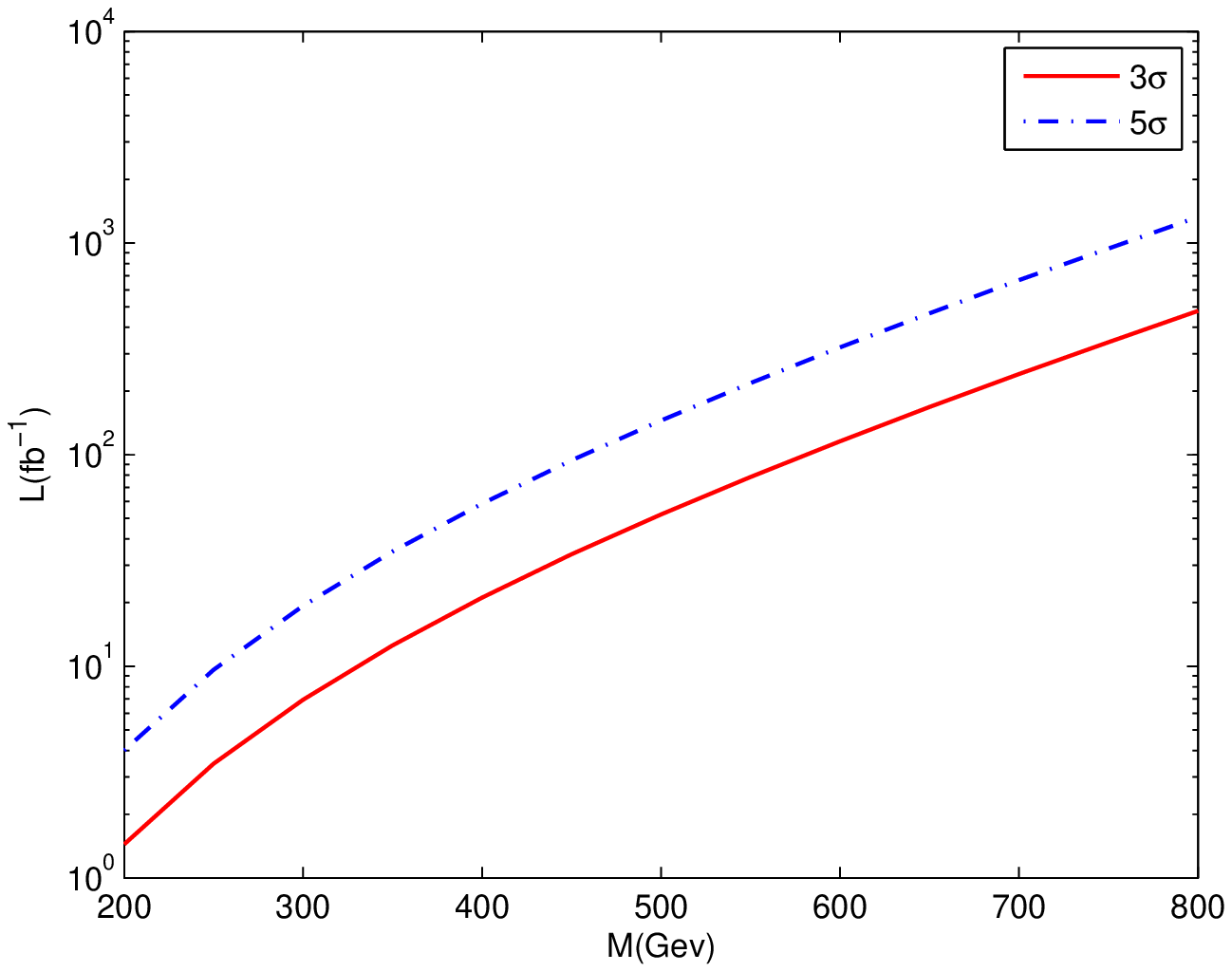}
\includegraphics[width=0.45\textwidth]{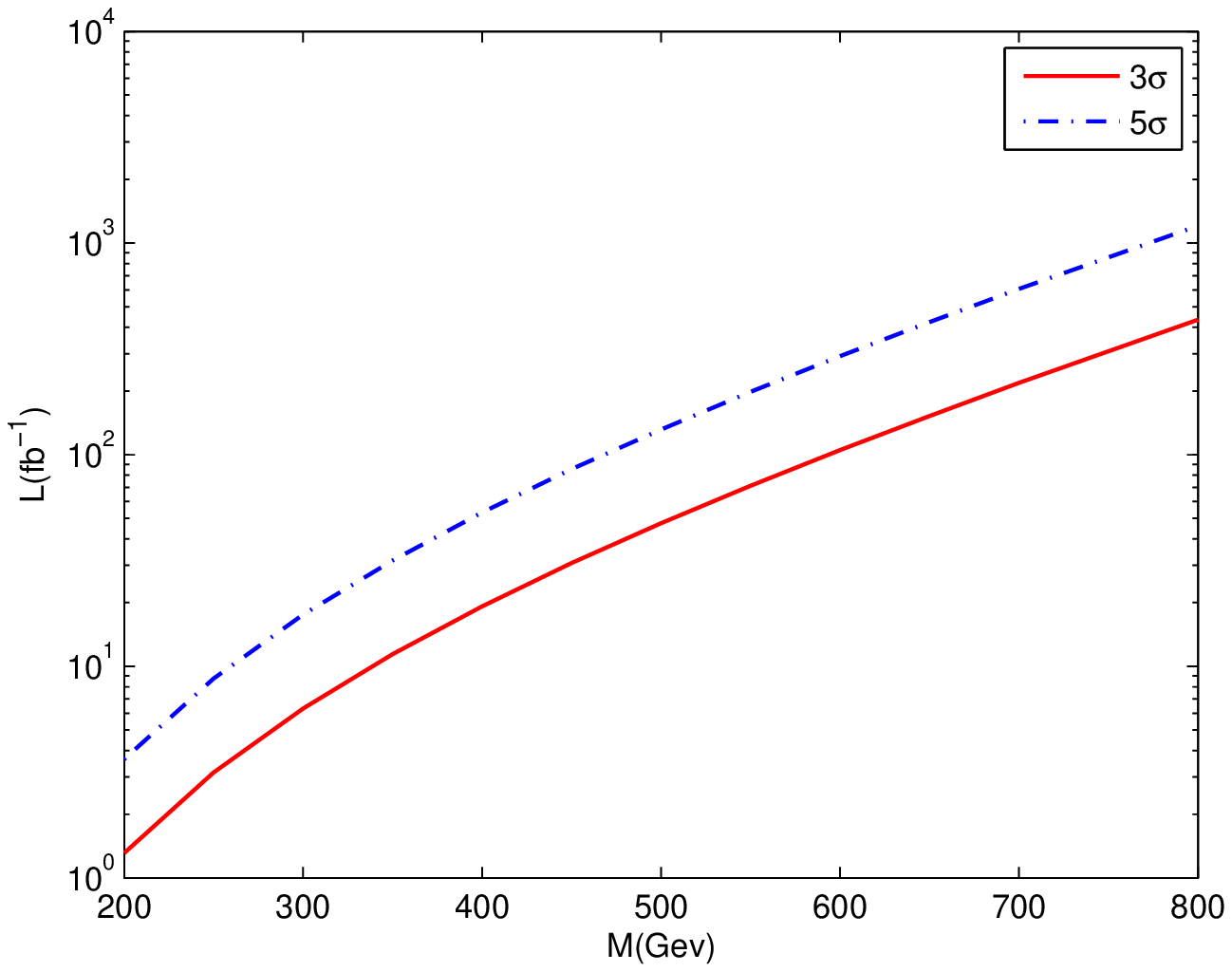}
\includegraphics[width=0.45\textwidth]{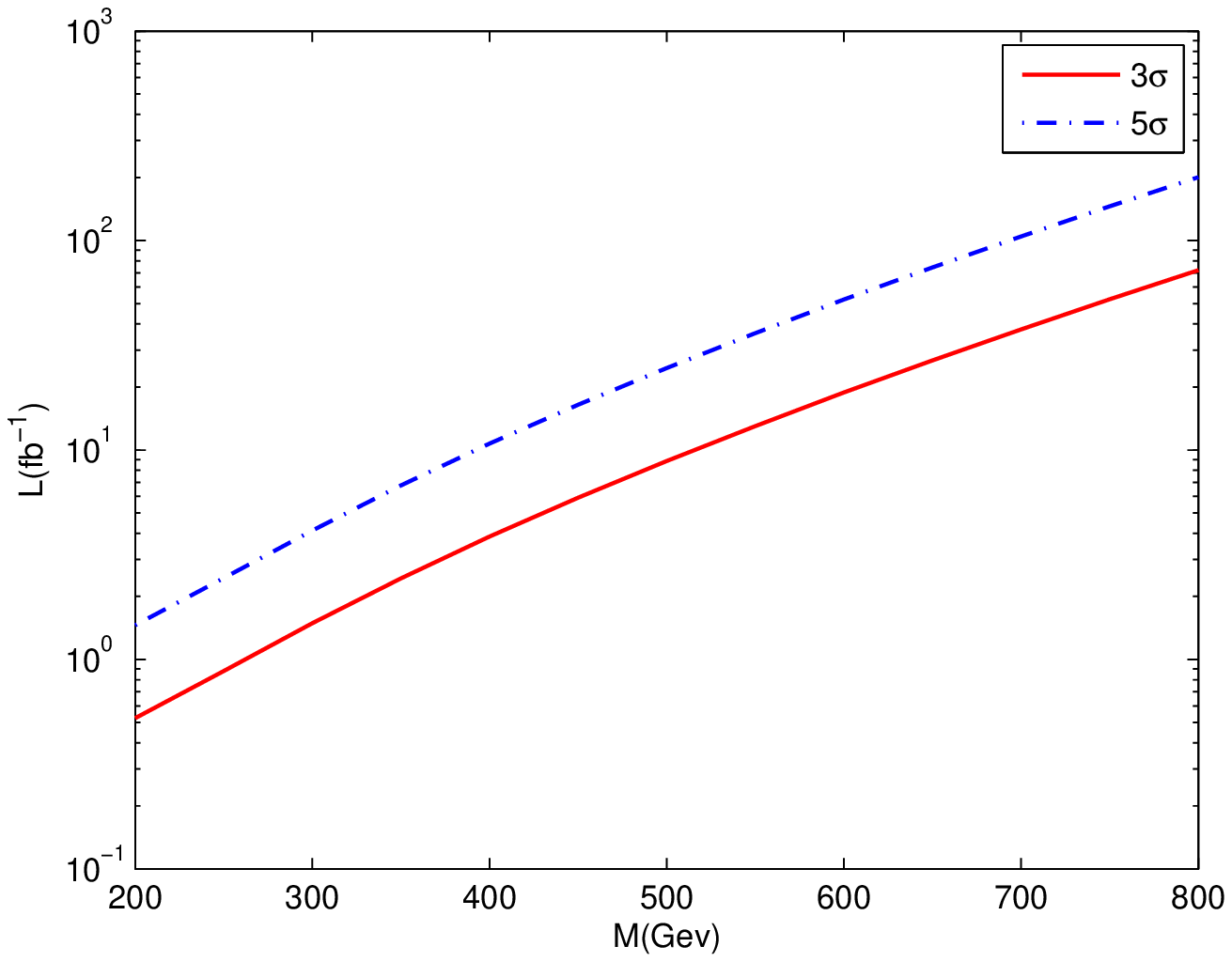}
\caption{The needed luminosity to observe different mass doublet leptons for the channels A (top-left), B (top-right), C (bottom-left) and D (bottom-right) for 3 $\sigma$ and 5 $\sigma$ significances at the 14 TeV LHC}\label{figure50-2}
\end{center}
\end{figure}

In Figure~\ref{figure50-1} (top-left panel) we also show the effective cross section for his channel after all the cuts.
We see that the cross sections are smaller than in the triplet case, because of the overall smaller production cross section.
Also, the contribution of the associated production channel, $X^{++} X^-$, is larger than for the triplet, because of the larger decay rate of $X^-$ into a $Z$ boson.
Finally, in Figure~\ref{figure50-2} (top-left panel) we plot the integrated luminosity needed at the LHC for a 5$\sigma$ discovery and 3$\sigma$ exclusion as a function of the mass of the exotic leptons.

\subsection{Singly charged lepton production: channel B, C and D}

In this section, we will discuss the channels which provide special signals to explore the presence of a singly-charged exotic lepton. The decay modes of singly charged lepton $X^{\pm}$ in the doublet model are different from the triplet model. They only decay to $Z$ and $H$ bosons without the $W$ channel, as showed in the Fig.\ref{figure-Doubletdecay}. These two decay modes can offer new potentially interesting channels and have larger branching fractions compared with the triplet leptons. We consider the decay to $Z$ boson and SM leptons at first, whose branching ratio is about $50\%-70\%$, much larger than triplet, see Fig.~\ref{figure-Doubletdecay}. Then we let the $Z$ boson decay to lepton pairs (channel B) and $b\bar{b}$ (channel C). Finally, we consider the decay into a Higgs boson plus a lepton, with subsequent decay of the Higgs into a $b\bar{b}$ pair (channel D).
For the other exotic lepton at production, which can be a doubly or singly charged one, we consider decays into a charged lepton plus two jets, originating from a $W$ in the case of a doubly-charged partner and a $Z$ for the singly charged.
The $b$ quarks are identified by requiring two $b$-tagged jets (we consider here a 70\% efficiency for the $b$-tag).
We therefore apply the same cut as in channel A for the $jj$ invariant mass, with an additional veto on missing energy $E\!\!\!\slash_T < 25 {\,\rm GeV}$.
In order to reconstruct the $Z$ or Higgs in the three channels, we impose the following additional cuts:
\begin{eqnarray}
\mbox{B)} &\Rightarrow & | M(\ell^+\ell^-) - M_Z |< 5 \,\rm GeV; \\
\mbox{C)} &\Rightarrow &| M(b \bar{b}) - M_Z | < 20 \,\rm GeV, \quad | M(\ell^+\ell^-) - M_Z | > 5 \,\rm GeV; \\
\mbox{C)} &\Rightarrow &| M(b \bar{b}) - M_h | < 20 \,\rm GeV, \quad | M(\ell^+\ell^-) - M_Z | > 5 \,\rm GeV.
\end{eqnarray}
Finally, we reconstruct the appropriate invariant masses of the two exotic leptons, and require that their values fall within $20$ GeV.
Note that in the case where more than one combination can be done, like for the cut on $M(\ell^+\ell^-)$ in channel B, we always pick the combination which yields the closer value to the bound.

For the three channels, we observed that the background is very suppressed, and it becomes negligible if we count only the events around the reconstructed invariant mass of the exotic leptons.
The final results of our simulation are shown in Figure~\ref{figure50-1}, where we show the effective cross sections after the cuts in the channels A--D.
In Figure~\ref{figure50-2} we also show the required luminosity at LHC14 for a 5$\sigma$ discovery and 3$\sigma$ exclusion as a function of the mass.
We see that, due to the absence of the decays into a $W$ for the singly charged lepton, these channels are much more important for the doublet than for the triplet case~\cite{Ma:2013tda}.
In fact, the channel D, which targets the Higgs decays, gives a sensitivity which is nearly as important as the channel A for the doublet model, and it becomes the golden channel for the discovery of this kind of exotic leptons.

\subsection{Singly charged lepton detection via $X^{+}X^{-}$ pair production}

The $X^{+}X^{-}$ pair production can also provide special final states as, in the doublet model, the singly charged pair always decays into two charged leptons and two neutral bosons ($Z$ or Higgs). All the neutral bosons can be detected via $b\bar{b}$ decays, as in channel E. We thus consider the following processes:
\begin{eqnarray}
pp \rightarrow X^{+}X^{-}  \rightarrow \left\{ \begin{array}{c}
\ell^+ Z\ell^- Z \\
\ell^+ H\ell^- H \\
\ell^+ Z\ell^- H  \end{array} \right\} \rightarrow \ell^+ \ell^- b\bar{b}b\bar{b}\,.
\end{eqnarray}

The signal final states are two $b\bar{b} $ pairs from $Z$ or Higgs boson decays and two leptons without missing energy. While the $X^{+}X^{-}$ pair production rate is small for triplet leptons, for the lepton doublet model it is enhanced due to the larger pair production cross section and larger branching fractions to two neutral bosons. Thus we expect this signal to be more relevant in this case than for triplets. Of course, the final detectable signal is suppressed by b-tagging efficiency if we require to tag all the 4 b-jets in the final state. After imposing the same basic acceptance cuts, neutral bosons mass cuts and exotic lepton mass matching cut as for the above channels, the background is negligible.

\begin{figure}
\begin{center}
\includegraphics[width=0.45\textwidth]{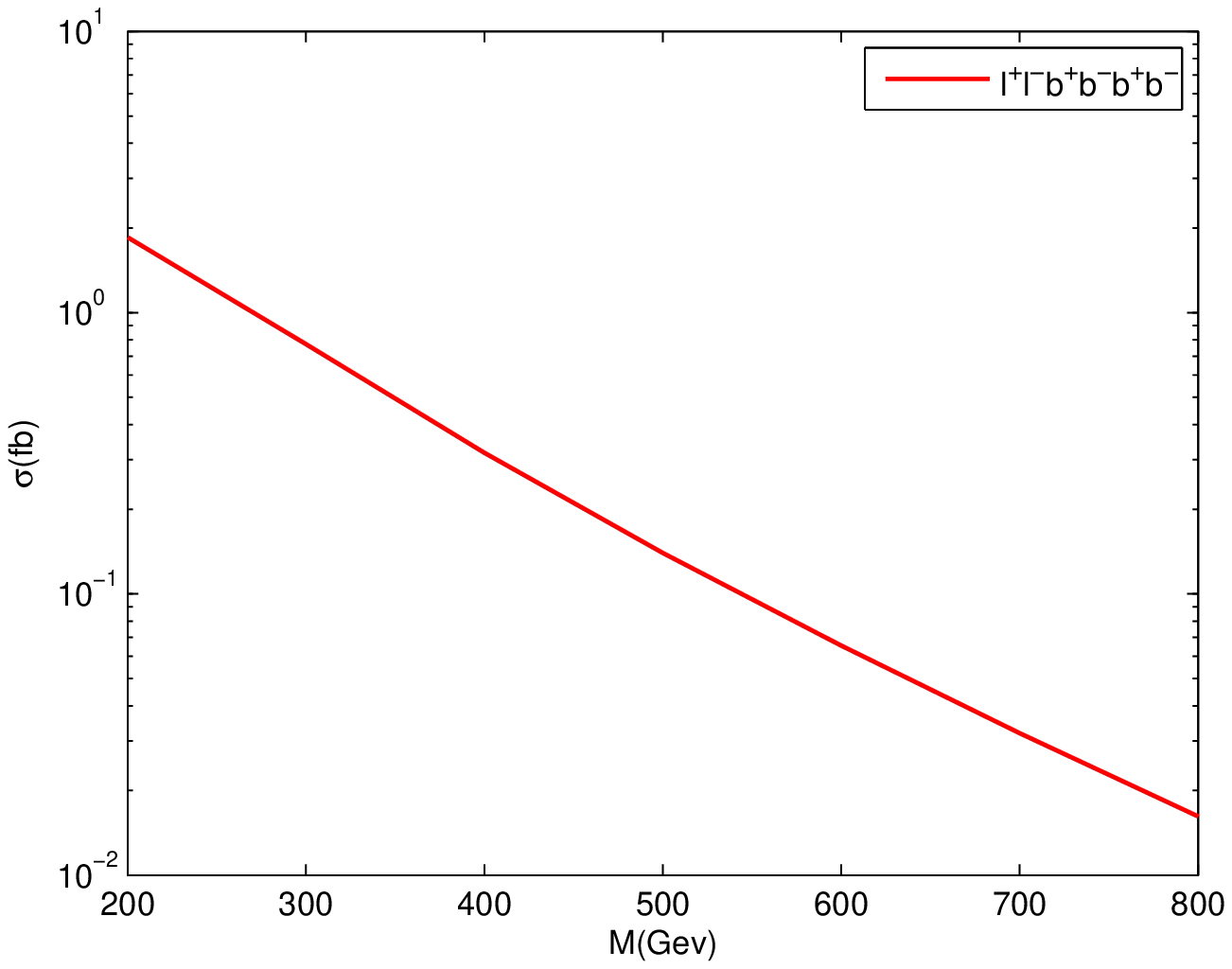}
\includegraphics[width=0.45\textwidth]{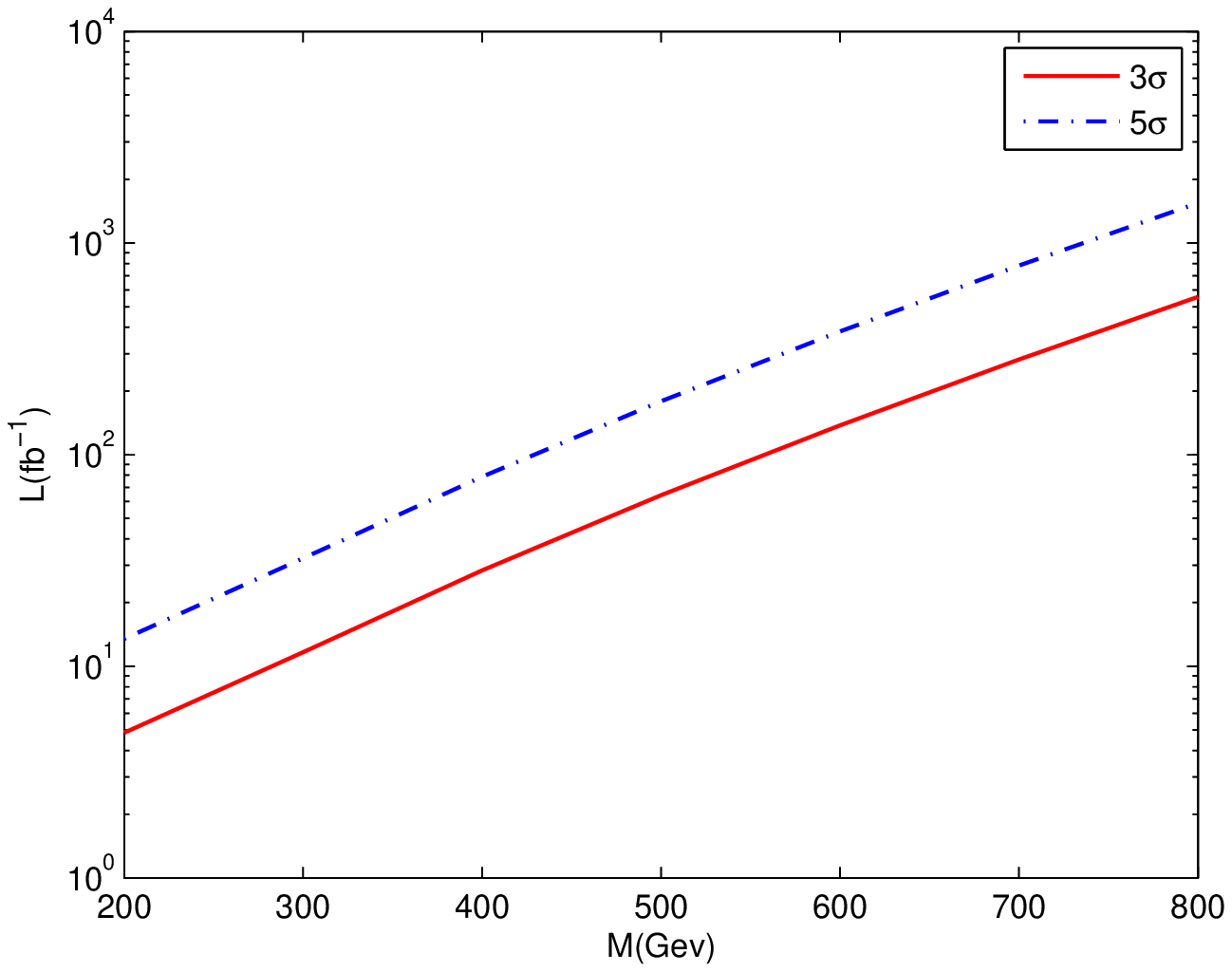}\\
\caption{Left: final lepton doublet signal production rate for $\ell^+\ell^-  b\bar{b}b\bar{b}$ channel at 14 TeV LHC. Right: needed luminosity to observe different mass doublet leptons via $\ell^+\ell^-   b\bar{b}b\bar{b} )$ channel for 3 $\sigma$ and 5 $\sigma$ significances at the 14 TeV LHC.}\label{figure-cs4b}
\end{center}
\end{figure}

The results of our simulation are shown in Fig.~\ref{figure-cs4b}: the effective cross section after cuts and the needed integrated luminosity to observe the doublet leptons $X$ as a function of the mass.
We can see that this channel is still subleading compared to the channels A and D.

\subsection{Lepton flavor violating signature of the doublet model at LHC }

We conclude this section with another important channel, which is however only relevant when the exotic leptons couple significantly to both electrons and muons.
In fact, the presence of a pair of leptons of different flavor allows to strongly reduce the standard model background.
The other advantage of this channel is that we can allow the bosons ($W$, $Z$ or Higgs) from the exotic lepton decay to decay hadronically, thus gaining signal from the larger hadronic branching ratios.
In the standard model, a pair of different flavor leptons can originate from the production of two $W$'s plus jets ($W^+ W^- jj jj$). However, this kind of background events always contain missing energy in the form of neutrinos.
A veto on missing energy $ E\!\!\!\slash_T < 25 {\,\rm GeV}$ can therefore reduce the background to reasonable levels.

In the following analysis, we will continue to assume that the mixing parameters with three generations are the same.
All of the exotic leptons pair production and associated production processes can provide the lepton flavor changing signature:
\begin{equation}
pp\rightarrow X^{--}X^{++} \rightarrow e^-  W^{-}\mu^+W^{+} +h.c. \rightarrow e^- \mu^+ j j jj+h.c.
  \label{31}
\end{equation}
\begin{equation}
pp\rightarrow X^{--}X^{+}+h.c. \rightarrow   e^-  W^{-}\mu^+Z(H) + h.c.  \rightarrow e^- \mu^+ j j jj+h.c.
  \label{32}
\end{equation}
\begin{equation}
pp\rightarrow X^{-}X^{+}  \rightarrow e^-\mu^+ZZ(ZH,HH) +h.c. \rightarrow e^- \mu^+ j j jj+h.c.
  \label{33}
\end{equation}
We only considered the hadronic decay mode of the gauge or Higgs boson in these signal processes,
and we do not require any $b$-tagging on the jets from the Higgs decay, although the light Higgs (125 GeV) boson decays most often to $b\bar{b}$.

The real rate of this signal depends on the couplings of the exotic leptons to electrons and muons, and our results can be generalized rescaling the signal events by a factor
\begin{equation}
\frac{9}{4} (1-\zeta_\tau)^2 \frac{4  |v_e|^2 |v_\mu|^2}{(|v_e|^2 + |v_\mu|^2)^2} < \frac{9}{4}\,.
\end{equation}

\begin{figure}
\begin{center}
\includegraphics[width=0.45\textwidth]{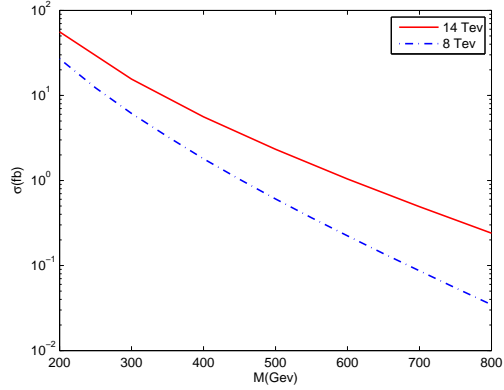}\\
\caption{The final cross sections for the lepton flavor violating signature at 8TeV and 14TeV LHC}\label{figure-flavor-cs}
\end{center}
\end{figure}

We require the presence of exactly 4 jets, with the same basic kinematic acceptance cuts as in the previous analyses (see Eq.(\ref{3})).
We also impose the $W/Z/H$ mass cuts on the jets as before to ensure that they come from gauge or Higgs boson decays, and we label the pairs whose invariant mass is close to any of the boson masses. Then we impose the mass matching cuts
\begin{equation}
| M(e jj) - M( \mu jj ) | < 20 \,\rm GeV.
\end{equation}
The two mass cuts can suppress the background to be neglected and to not affect the signal. The effective cross sections for the lepton flavor violating signature after the cuts at 8 and 14 TeV are given in the Fig.~\ref{figure-flavor-cs}.

\begin{figure}
\begin{center}
\includegraphics[width=0.45\textwidth]{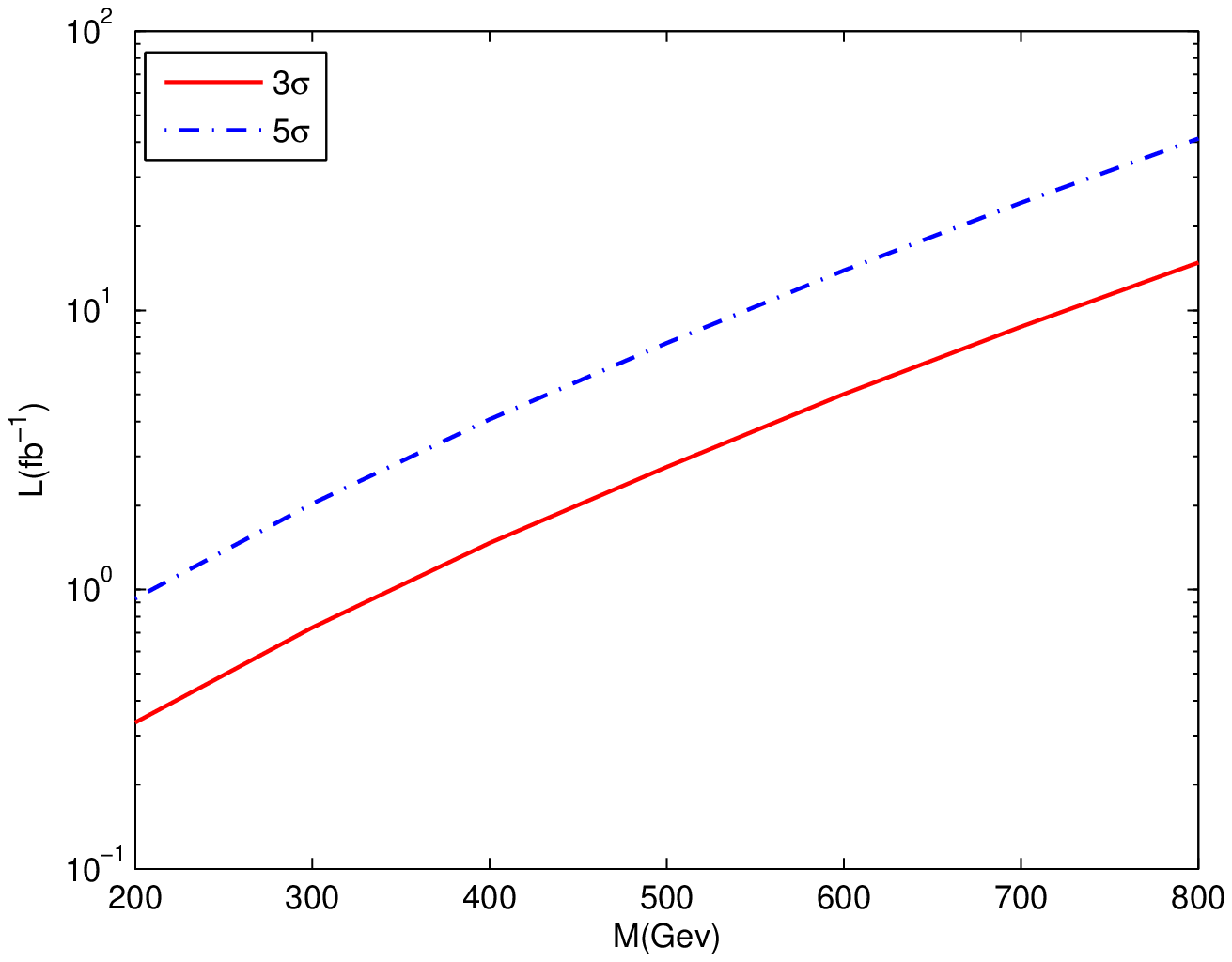}
\includegraphics[width=0.45\textwidth]{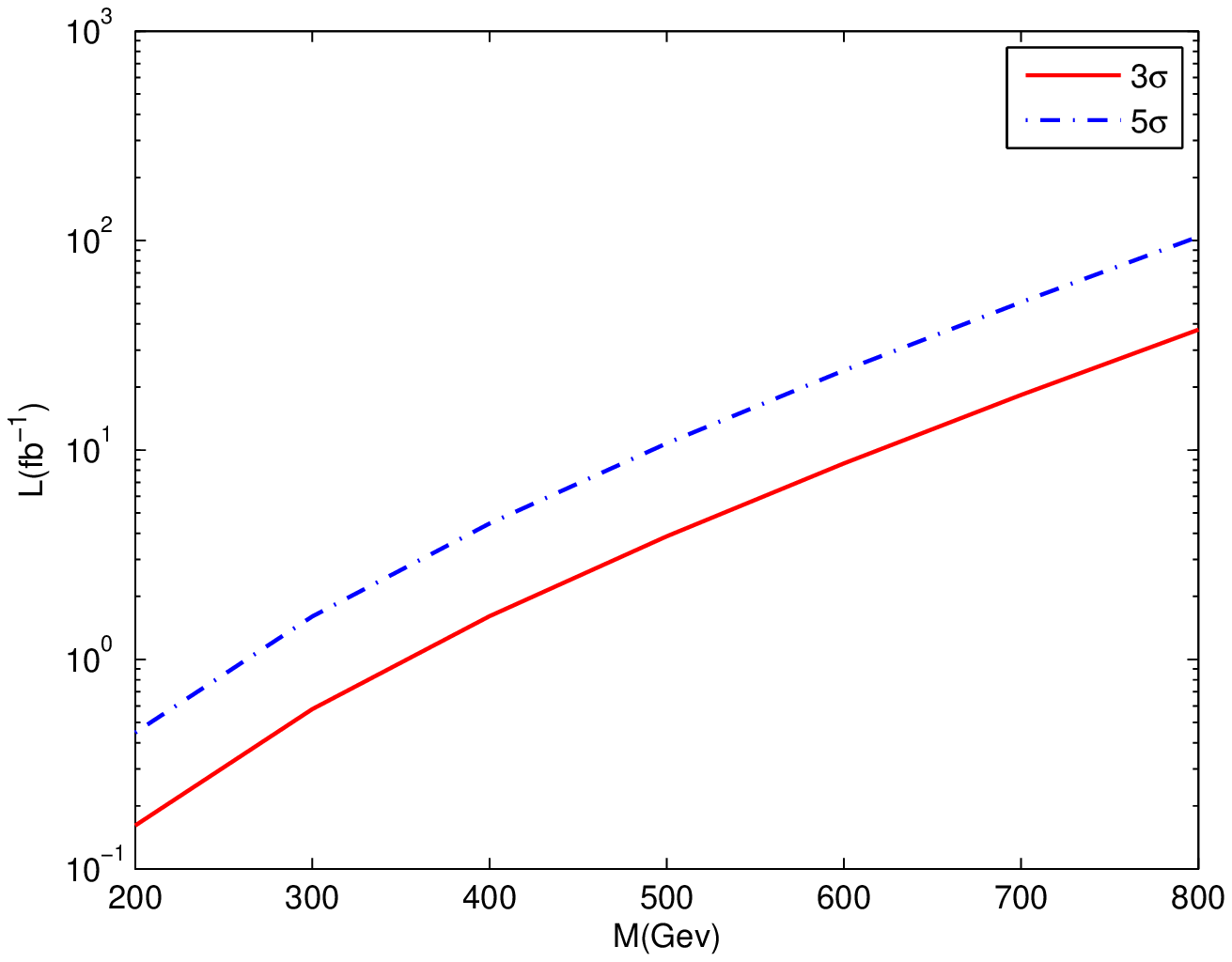}\\
\caption{Needed luminosity to observe different mass doublet leptons via lepton flavor violating processes at the 8 TeV (left) and 14 TeV (right) LHC.}\label{figure5555-2}
\end{center}
\end{figure}

The cross sections are almost the same as for the triplet model. We can see from the Fig.~\ref{figure5555-2} that the 20/fb data at 8 TeV LHC have the potential to discover the lepton doublet model for small masses ($M_X<400$GeV).
 We plotted the needed luminosity for discovery/exclusion at 14 TeV in the right panel of Fig.~\ref{figure5555-2}, the potential to observe the lepton flavor violating channel is better than all the above channels in the flavor structure of the couplings we chose.

\section{Distinguishing doublet from triplet exotic leptons.}

From the previous discussion, we can see that the production rates of various signatures are different for the triplet or doublet leptons. We can therefore distinguish the two models by measuring and comparing the production rates.
In particular, we have shown that in the doublet model case D ($\ell^- \ell^- H(b\bar{b}) jj$) offers similar rates as the same-sign di-lepton channel, while it is subleading in the triplet case.

\begin{figure}
\begin{center}
\includegraphics[width=0.6\textwidth]{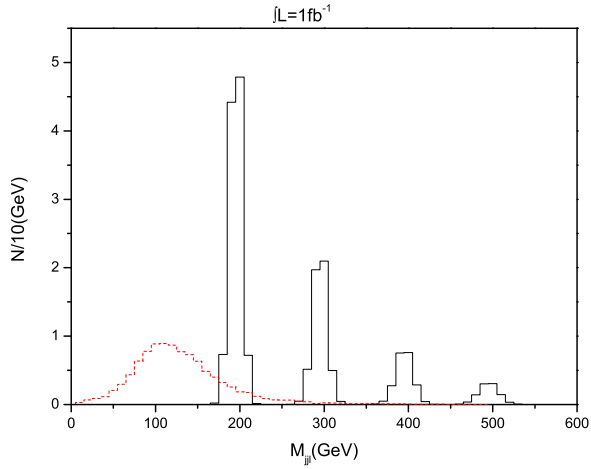}\\
\caption{Invariant mass $M(\ell jj)$ distribution curves of exotic leptons signal resonances in the $X^\pm X^0\to \ell^\pm \nu jj  b\bar{b}$ channel with the background.}\label{figure-X1X0IM}
\end{center}
\end{figure}

\begin{figure}
\begin{center}
\includegraphics[width=0.45\textwidth]{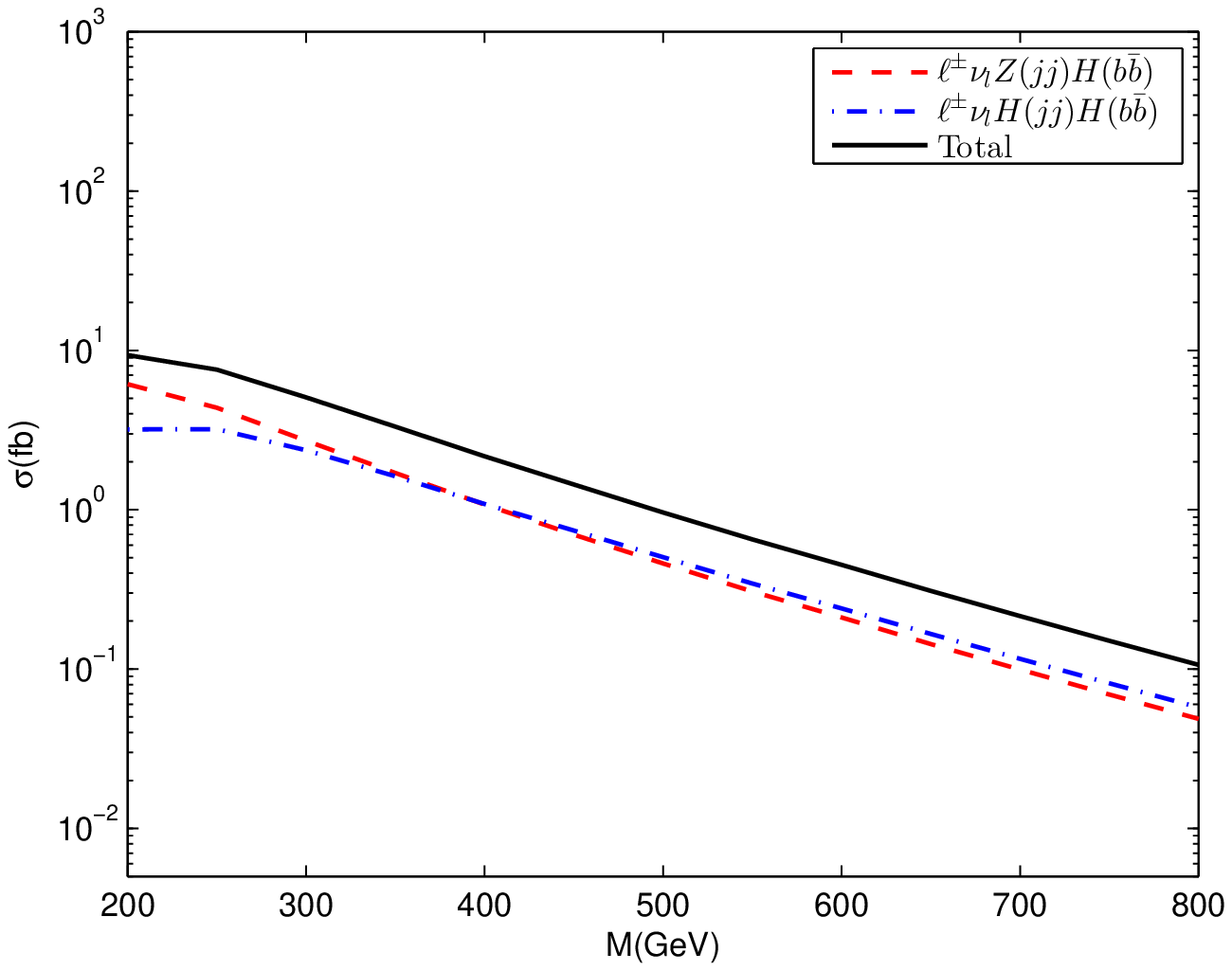}
\includegraphics[width=0.45\textwidth]{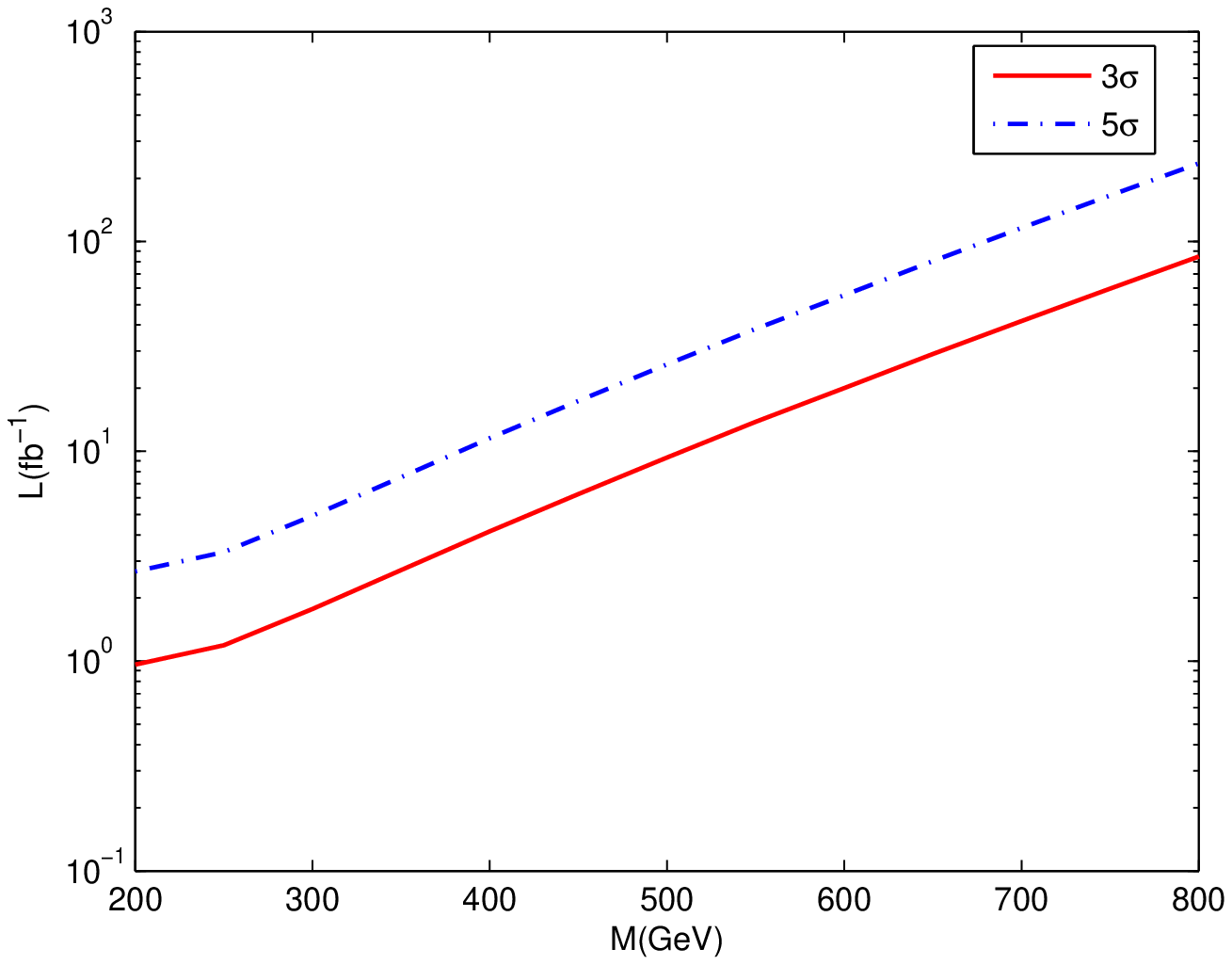}\\
\caption{Left: final lepton triplet signal production rate for $X^\pm X^0\to \ell^\pm \nu jj  b\bar{b}$ channel at 14 TeV LHC. Right: needed luminosity to observe different mass triplet leptons via $\ell^\pm \nu jj  b\bar{b})$ channel for 3 $\sigma$ and 5 $\sigma$ significances at the 14 TeV LHC.}\label{figure-X1X0CSLU}
\end{center}
\end{figure}

In addition to the difference between production rates of those common channels, the triplet contains exotic neutral leptons $X^0$ which are absent in the lepton doublet.  The neutral leptons $X^0$ can be produced associated with $X^\pm$ via Drell-Yan process at LHC. We didn't consider this process in our previous paper~\cite{Ma:2013tda} because the $X^\pm X^0$ channel has no advantage in discovering the exotic leptons with respect to the other channels. However, detecting this associated production process $pp\to X^\pm X^0$ may be useful to prove if the exotic leptons belong to a triplet or a doublet. The neutral lepton $X^0$ can decay to a neutrino and a neutral $Z$ or Higgs boson, the branching fractions are given in the Fig.3 of Ref.~\cite{Ma:2013tda}. The $X^\pm$ also can decay to a neutrino and a $W$ boson, but we can not reconstruct the invariant mass of $X$ leptons in this decay mode. Therefore, we only consider the  $X^\pm$ lepton decay to a charged lepton and a neutral $Z$ or Higgs boson. So the process is
\begin{equation}
pp\to X^\pm X^0 \to \ell \nu ZZ(HZ,HH) .
\end{equation}
In order to avoid the large QCD background, we require at least one of the two neutral boson to decay to a pair of b-jets or a lepton pair in the signal.
After considering the branching fractions, the dominant signature processes are
\begin{eqnarray}
pp\to X^\pm X^0 \to \ell \nu Z(jj) H(b\bar {b}) ;\\
pp\to X^\pm X^0 \to \ell \nu H(jj) H(b\bar {b}) .
\end{eqnarray}
The other possible decay channels as $Z\to \ell^+\ell^-$ and $Z\to b\bar{b}$ only contribute small production rates. So we fix the signal as $\ell \nu jj b\bar {b}$ in the final states, where the b-jet pair is from Higgs boson decay and the other two jets are from $Z$ or Higgs boson decay. Then the invariant mass cuts are imposed as follows:
\begin{eqnarray}
 & | M(b \bar{b}) - M_h | < 20 \,{\rm  GeV}, & \\
 & M_Z-20 \,{\rm GeV} <M(jj) < M_h + 20\,{\rm GeV}. &
\end{eqnarray}
However, even after the mass cuts, the background is still covering the signal. The dominant background is $jjb\bar{b}W(\ell \nu)$. In order to suppress the background in which the lepton is from $W$ boson decay, we imposed the transverse mass $M_T$~\cite{Summers:1996yc} cut as
\begin{equation}
M_T(\ell \nu) > M_W .
\end{equation}
This cut can rapidly reduce the background while affect the signal slightly. Although there is missing energy in the final states and the $X$ invariant mass matching cut is not available, we still can reconstruct the $X^\pm$ mass as $M(\ell jj)$. In the right panel of Fig.~\ref{figure-X1X0IM}, we plotted the distribution of the invariant mass $M(\ell jj)$ for exotic lepton signals at different masses and the background. The surviving SM background events are distributed at the low invariant mass region where the resonance width is narrow, thus the background is negligible if we only consider the events within the signal resonances.
After considering all the cuts and decay branching fractions, the effective signal cross section is still sizable when the mass of lepton triplet is not larger than 700 GeV as showed in the left panel of Fig.\ref{figure-X1X0CSLU}. The right panel of Fig.\ref{figure-X1X0CSLU} shows the needed integrated luminosity to observe the triplet leptons $X^0$ as a function of the mass.  Detecting this channel, unique to the triplet model, can thus disentangle the doublet and triplet models.

 Another method to distinguish doublet from triplet is by studying the angle distribution of the $X$ decay final states: in fact, a difference between the two models is the chirality of the coupling of the $X^{--}$ to the $W^-$ boson. This same method has been used to determine the top quark chiral couplings with $W$ boson and with an extra gauge boson $W'$~\cite{Kane:1991bg,Gopalakrishna:2010xm}. Both the triplet and doublet leptons have non-chiral couplings to the $W$ and $Z$ in the pair production, thus the orientation of the spin of the produced $X$ is the same and there is no preferred direction for the lepton from the $X$ decay. However, the triplet $X^{--}$ dominantly decays to left-handed $\ell^-$ with a left-handed or longitudinal $W$, while opposite chiralities are produced by the doublet. Unfortunately, the spin (or chirality) of leptons cannot be determined at the LHC, however the chirality of the $W$ boson will be reflected in the angular distribution of the lepton from the $W$-decays.

We can analytically derive the partial width of the following leptonic decay of doubly-charged lepton
   \begin{equation}
 X^{--}\rightarrow \ell^- W^{-}\rightarrow \ell^- \ell^- \bar{\nu_{\ell}}
  \label{34}
\end{equation}
We firstly consider the direction and energy of the first lepton from $X^{--}$ decay in the rest frame of $X^{--}$.
The lepton's energy is given as:
\begin{equation}
E_{1}=\frac{M^2-m_{W}^2}{2M},
\end{equation}
where, $M$ is the mass of $X^{--}$. Meanwhile, the energy of $W$ in $X^{--}$ rest frame is
\begin{equation}
 E_{W}=\frac{M^2+m_{W}^2}{2M}
 \end{equation}
 Then, the $W$ will decay in to another lepton and a neutrino, the energy of the second lepton is also given as $M_W/2$ in the  $W$  rest frame. Of course, the outgoing  direction of the second lepton is arbitrary. We define the angle between  the second lepton's direction in the  $W$  rest frame and the $W$ boson's direction in the rest frame of $X^{--}$ as $\theta$. Then, we can get the energy of the second lepton  as
\begin{equation}
E_{2}=\frac{E_1\cos\theta+E_W}{2},
\end{equation}
and the neutrino energy as
\begin{equation}
E_{3}=\frac{-E_1\cos\theta+E_W}{2}.
\end{equation}
Finally, the triplet $X^{--}$ decay partial width depends on $\cos \theta$ as
\begin{equation}
\left. \frac{d\Gamma}{d\cos \theta} \right|_{\rm triplet} \propto E_{3}(1+\cos\theta),
\end{equation}
for left-handed couplings
%
On the other hand, for the right-handed couplings in doublet model:
\begin{equation}
\left.  \frac{d\Gamma}{d\cos \theta} \right|_{\rm doublet} \propto E_{2}(1-\cos\theta),
\end{equation}



\begin{figure}
\begin{center}
\includegraphics[width=0.45\textwidth]{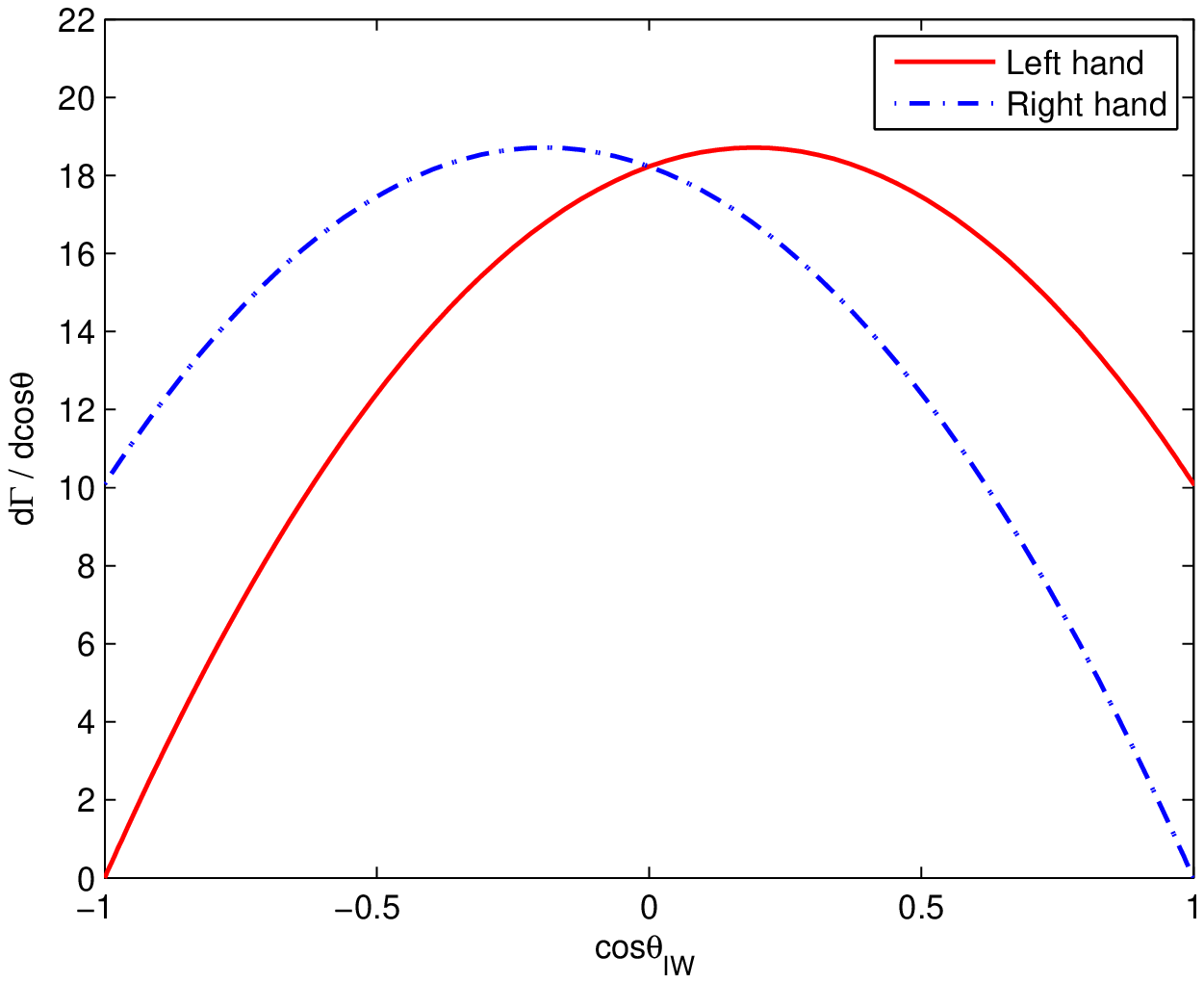}
\includegraphics[width=0.45\textwidth]{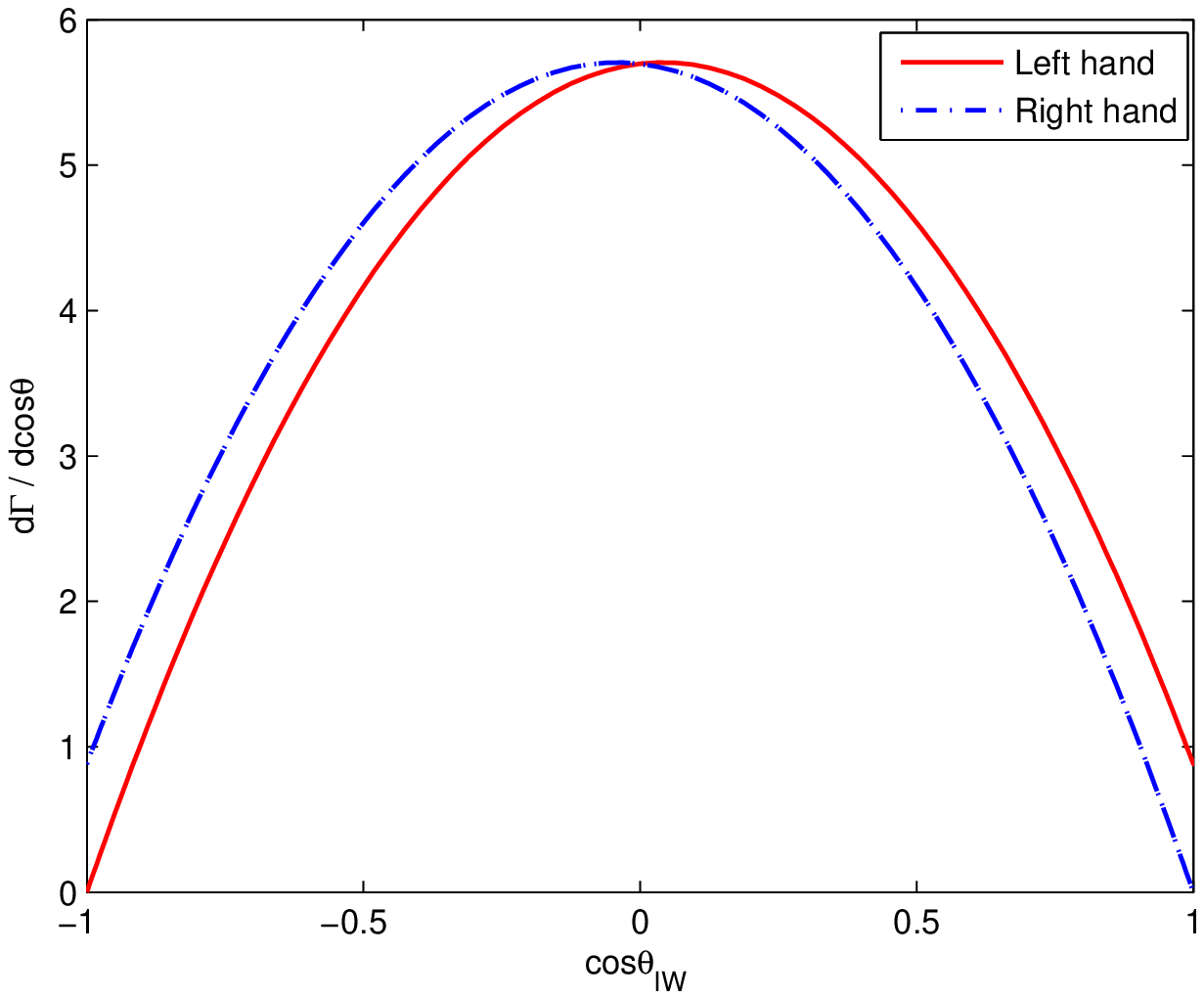}\\
\includegraphics[width=0.45\textwidth]{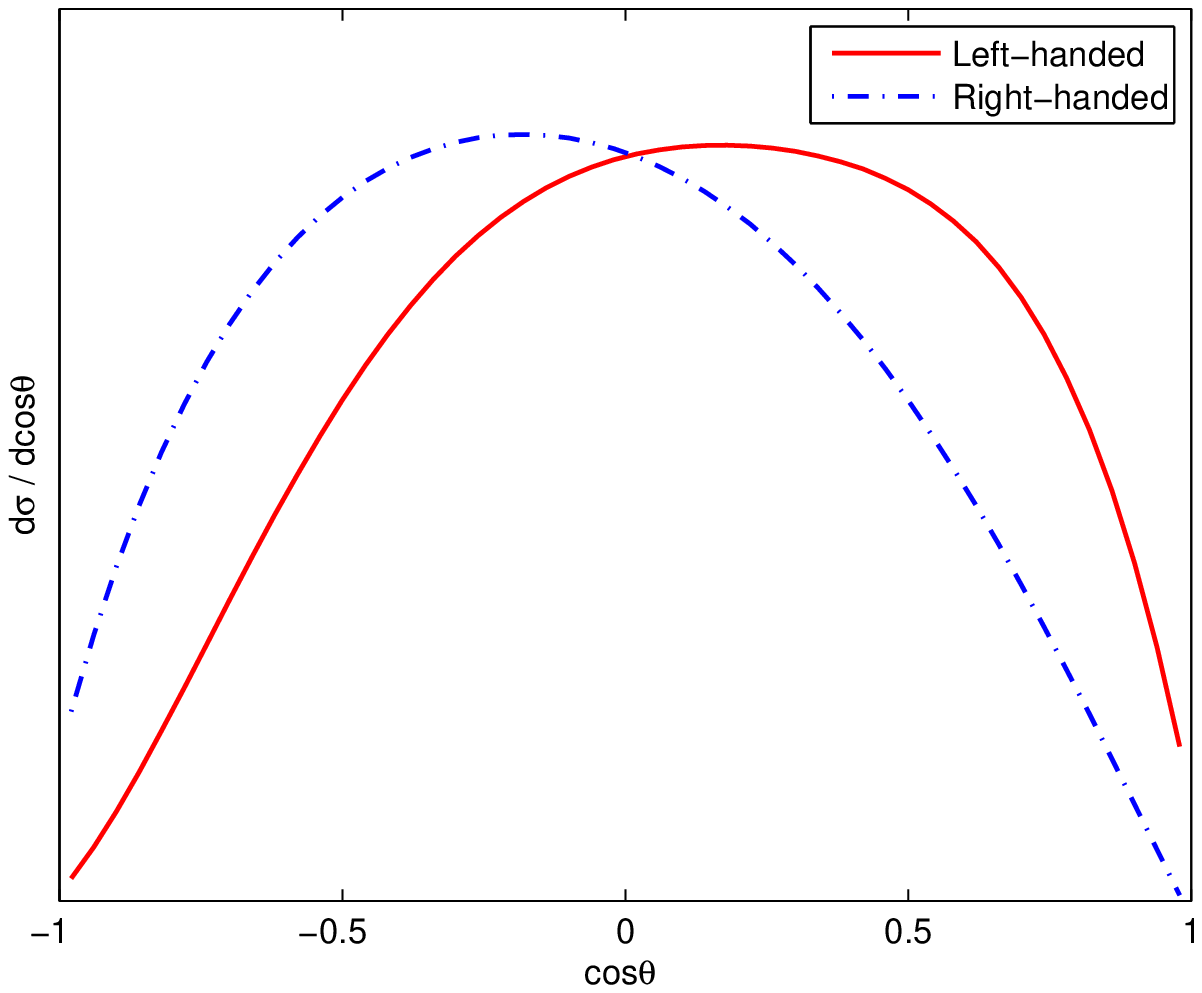}
\includegraphics[width=0.45\textwidth]{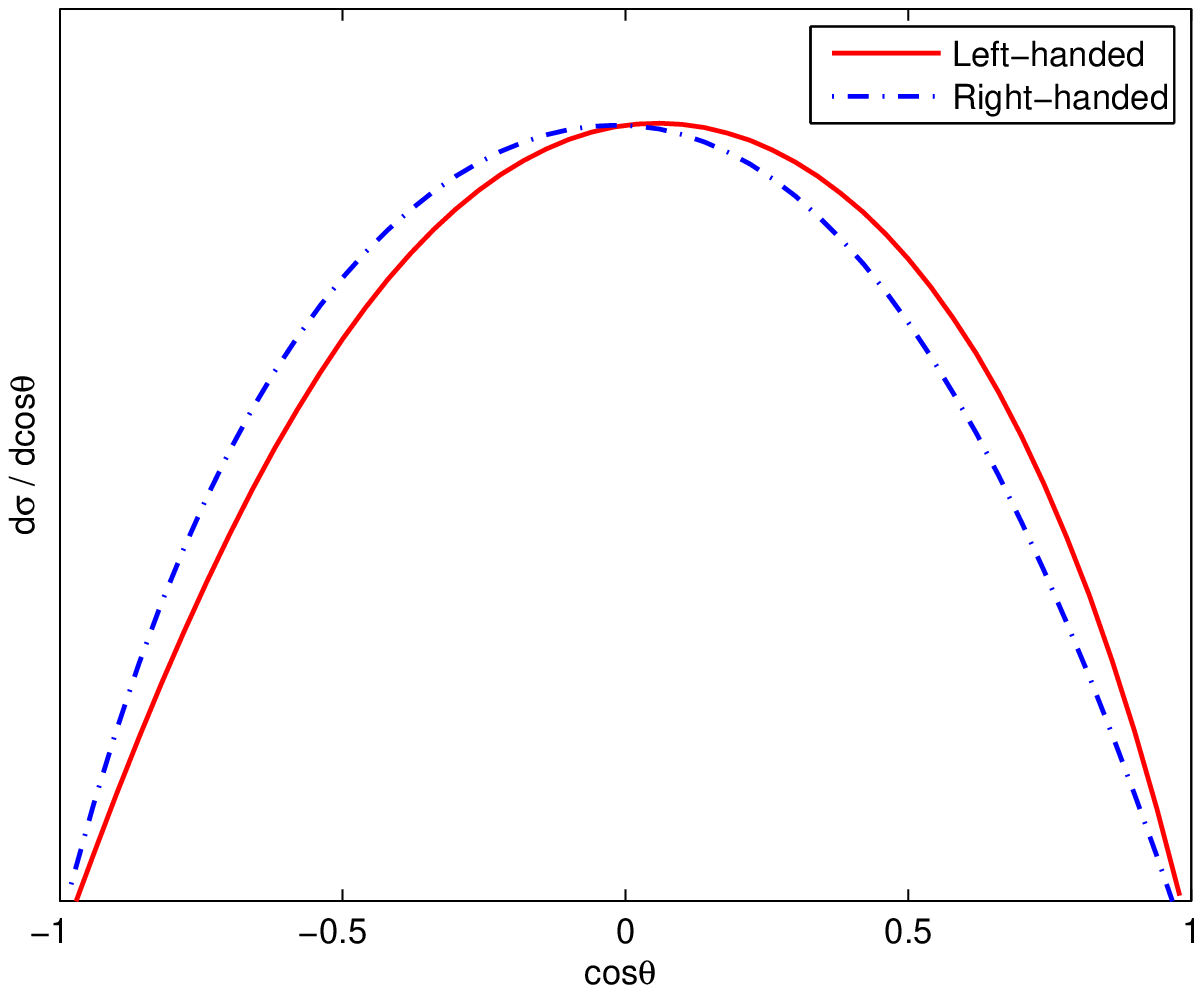}\\
\caption{ Normalized angular distribution in $ X^{--}$ decay with mass $M_{X}=200$ GeV (left) and $400$ GeV (right) for triplet left-handed mixing and doublet right-handed mixing. The plots on the top row correspond to events before the cuts, the bottom row after cuts.}\label{figure6-1}
\end{center}
\end{figure}

\begin{figure}
\begin{center}
\includegraphics[width=0.45\textwidth]{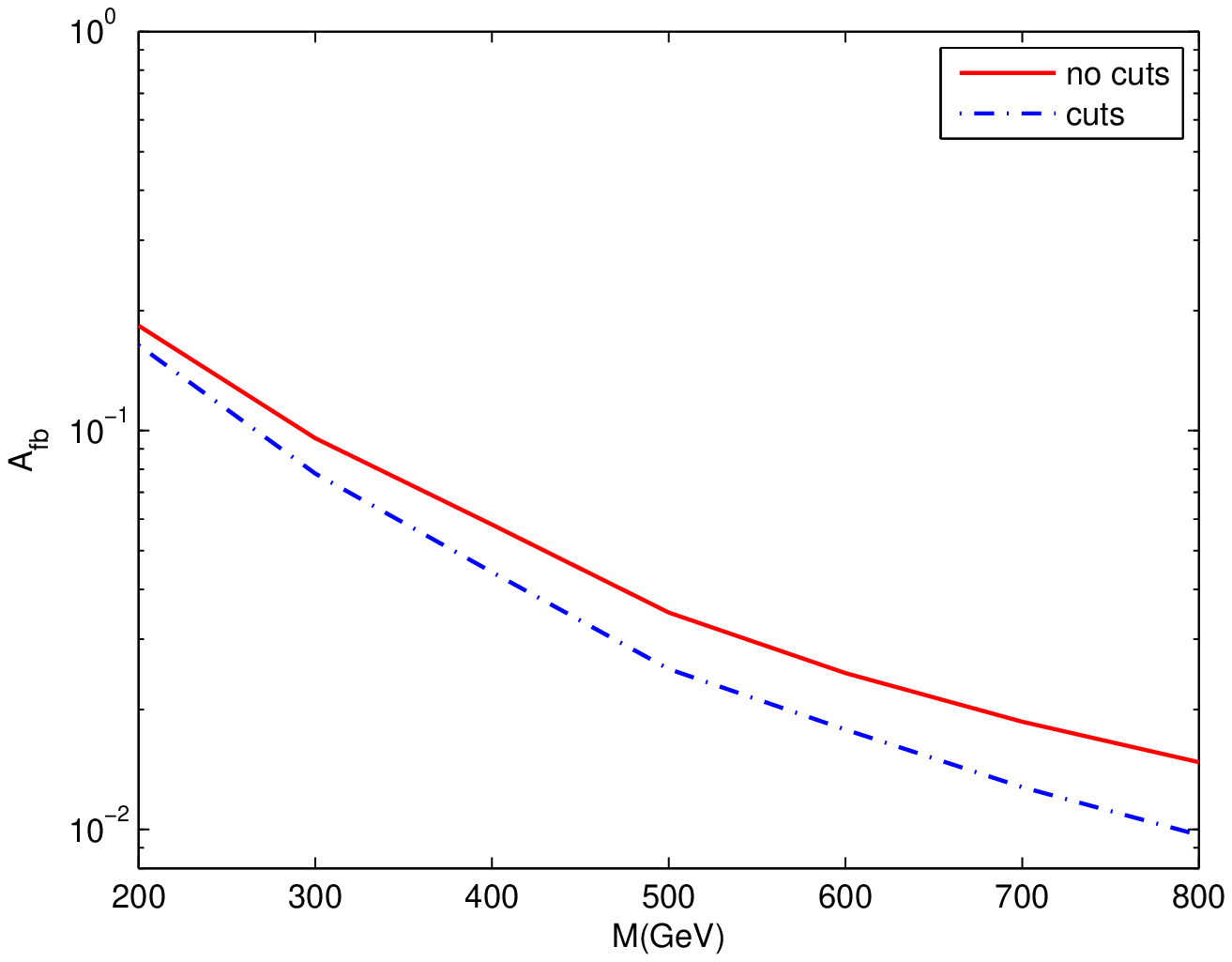}
\includegraphics[width=0.5\textwidth]{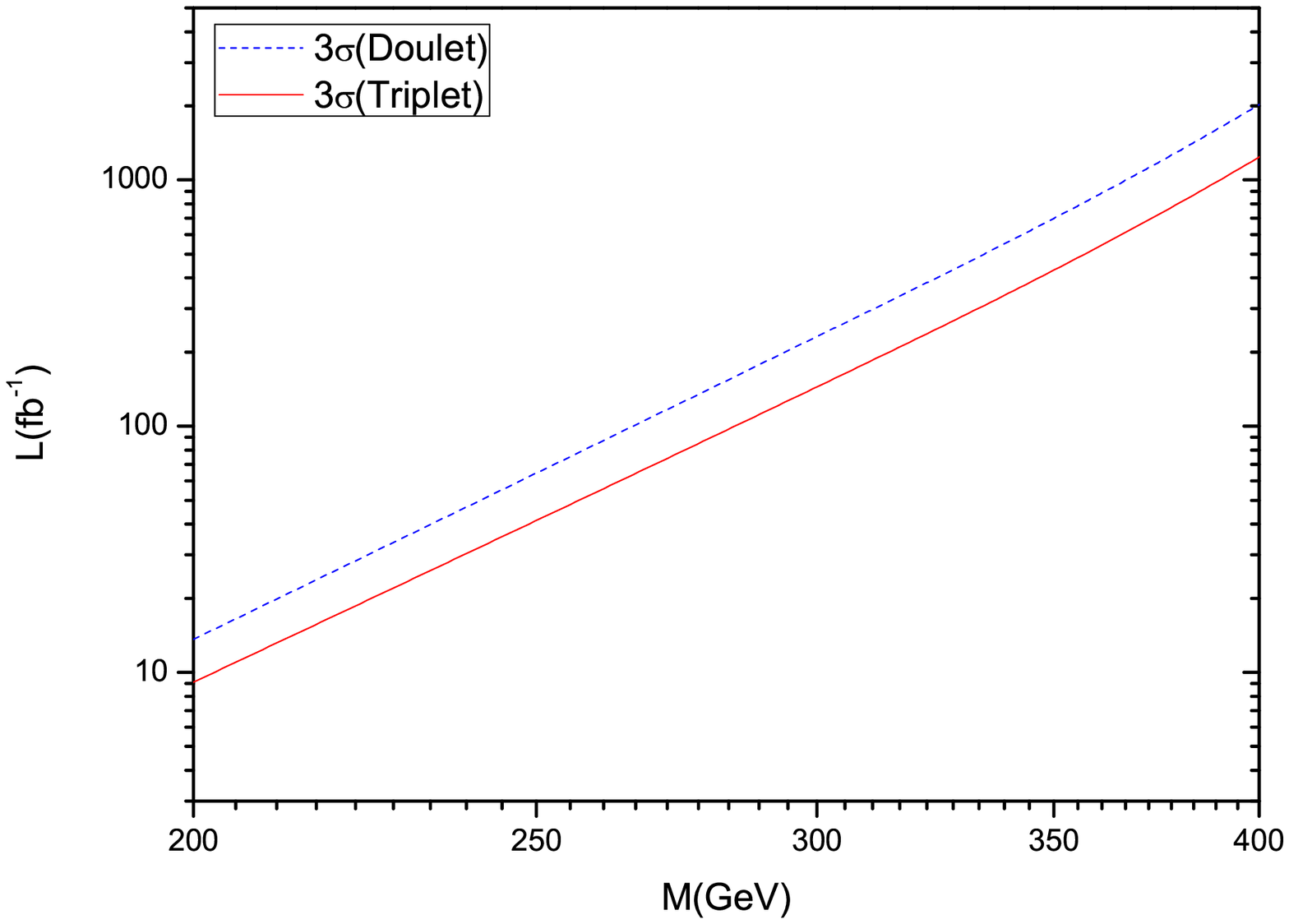}\\
\caption{Left: asymmetry $A_{fb}$ in the angular distribution before (solid) and after(dashed) cuts as a function of the exotic lepton mass. Right: needed luminosity to determine the chirality of the exotic lepton coupling with 3 $\sigma$ significance at the 14 TeV LHC.}\label{figure-asymmetry}
\end{center}
\end{figure}

We plot the normalized distribution of $cos\theta$ in $X^{--}$ decay processes in the Fig.~\ref{figure6-1} (top row) for $M_{X}=200$ GeV (left) and 400 GeV (right).
The distributions in actual experiments would not be so clear though, because of the effect of the kinematical cuts. We simulated the angular distributions after cuts and show the results in the bottom row of Fig.~\ref{figure6-1} for the same values of masses. After considering all the cuts, the asymmetry of the angular distribution is reduced as expected.
Nevertheless, we define a forward-backward asymmetry parameter
\begin{equation}
A_{fb}  =\frac{|N_f-N_b|}{N_f+N_b},
\end{equation}
where $N_f (N_b)$ is the event number with $\cos \theta>0$ ($\cos \theta<0$).
The asymmetry in the angular distribution decreases with increasing exotic lepton mass, as shown in the left panel of Fig.~\ref{figure-asymmetry}.
If we define the difference between forward and backward event number as the signal to determine the chirality of the exotic lepton's mixing to standard lepton, then $N_s=|N_f-N_b|$. Therefore the statistical significance can be defined as
\begin{equation}
s_{fb}  =\frac{|N_f-N_b|}{\sqrt{N_f+N_b}}=A_{fb}\sqrt{\sigma \cal{L}},
\end{equation}
where, $\sigma$ is the final total cross section and $\cal L$ is the integrated luminosity. The integrated luminosity needed to determine the chirality of the mixing to standard lepton is plotted in the right panel of Fig.~\ref{figure-asymmetry}: we can see that the LHC14 can distinguish the left-handed mixing of the lepton triplet from the right-handed mixing of the lepton doublet only when the mass less than 300 GeV.


\section{Conclusions}

In this work we considered the phenomenology at the LHC at 14 TeV of a lepton doublet which embeds a
doubly charged lepton. The exotic lepton doublet mixes with ordinary leptons via Yukawa couplings with the SM Higgs doublet. The only other case containing a doubly-charged lepton is a triplet. However, the doublet is different from the triplet case in that the mixing is dominantly in the right-handed sector is dominant as the mixing in the left-handed one is very small being suppressed by the small masses of the standard leptons. We analyzed the exotic doubly-charged lepton pair and associated production rates and their decay branching fractions as in the work of lepton triplet. The phenomenology at LHC of the lepton doublet is different from lepton triplet:
the best channel to search is not the tri-lepton final state $\ell^-\ell^+ jj \ell^- \bar{\nu}(\ell^+ \nu)$ which provides most signal in triplet case, but is the $\ell^-\ell^+ Z jj$ final state. We also analyzed all the other detectable channels as we considered in the triplet case, including the 4 b-jets channel of $X^+ X^-$ production and the lepton flavor violating signature of the doublet model. We find again that a search based on two different flavor leptons plus jets without missing energy is very effective in the case where significant mixing to both electrons and muons is present, and the lepton flavor violating signature may allow to pose significant bounds on the exotic doublet lepton mass already with the data at 8 TeV.

 In addition to the different rates, we explored other ways to distinguish the doublet from the triplet case.
We suggested the analysis of a neutral exotic lepton channel, which only exists in the triplet model.
We also find that the chirality of the exotic lepton mixing to standard leptons can be reflected in the angular distribution of the lepton from the $W$ decay, however it is experimentally challenging to measure such an effect unless for very light masses.

{\bf Acknowledgment:} The work of B.~Z. is supported by the
National Science Foundation of China under Grant No. 11075086 and
11135003. G.~C. would like to thank the Physics Department in Tsinghua University for the hospitality during the initial stages of this work, the visit being sponsored by a project of the France-China Particle Physics Laboratory (FCPPL) and an exchange program of the CNRS/NSFC. G.C. also acknowledges partial support from
the Labex-LIO (Lyon Institute of Origins) under grant ANR-10-LABX-66.

\end{document}